\documentclass{article}

\usepackage{arxiv}

\usepackage[utf8]{inputenc} %
\usepackage[T1]{fontenc}    %
\usepackage{url}            %
\usepackage{booktabs}       %
\usepackage{nicefrac}       %
\usepackage{microtype}      %
\usepackage{lipsum}
\usepackage{graphicx}
\usepackage{xspace}
\graphicspath{ {./images/} }

\usepackage[round]{natbib}

\usepackage{hyperref}       %
\hypersetup{colorlinks=true,citecolor={red},linkcolor={blue},linktocpage} 
\usepackage{bm}
\usepackage{bbm}
\usepackage{amsfonts}
\usepackage{amssymb} 
\usepackage{amsmath}
\usepackage{amsthm}
\usepackage{float}
\usepackage{subcaption}
\usepackage{xr} %
\usepackage{makecell}
\usepackage{algorithm}
\usepackage{algpseudocode}
\usepackage[titletoc]{appendix}

\theoremstyle{definition}
\newtheorem{definition}{Definition}[section]

\newcommand{\mS}{\mathcal{S}}

\newcommand{\bX}{\bm{X}}
\newcommand{\bx}{\bm{x}}
\newcommand{\by}{\bm{y}}
\newcommand{\bq}{\bm{q}}
\newcommand{\Lperm}{{L_{\text{perm}}}}
\newcommand{\Lrep}{{L_{\text{rep}}}}
\newcommand{\Labc}{{L_{\text{abc}}}}
\newcommand{\Xirr}{\bX_{\mathcal{S}_1}}
\newcommand{\Xrel}{\bX_{\mathcal{S}_0}}
\newcommand{\norm}[1]{\left\lVert#1\right\rVert}
\newcommand{\Local}{BART VIP-Local\xspace}
\newcommand{\Gse}{BART VIP-G.SE\xspace}
\newcommand{\Gmax}{BART VIP-G.Max\xspace}
\newcommand{\MI}{BART MI-Local\xspace}
\newcommand{\dart}{DART\xspace}
\newcommand{\abc}{ABC Bayesian forests\xspace}
\newcommand{\vipRank}{BART VIP Rank\xspace}
\newcommand{\vcMeasure}{DART VC-measure\xspace}

\usepackage{arydshln}

\allowdisplaybreaks

\title{Posterior Summarization for Variable Selection in Bayesian Tree Ensembles}

\author{
 Shengbin Ye \\
  Department of Statistics and Data Science\\
  Northwestern University\\
  Evanston, IL 60201 \\
  \texttt{sy21@northwestern.edu} \\
   \And
 Meng Li \\
  Department of Statistics\\
  Rice University\\
  Houston, TX 77005 \\
  \texttt{meng@rice.edu} \\
}

\begin{document}
\maketitle
\begin{abstract}
Variable selection remains a fundamental challenge in statistics, especially in nonparametric settings where model complexity can obscure interpretability. Bayesian tree ensembles, particularly the popular Bayesian additive regression trees (BART) and their rich variants, offer strong predictive performance with interpretable variable importance measures. We modularize variable selection with Bayesian tree ensembles into two components, the tree prior and the posterior summary, and show that, although typically framed as a modeling task, it often hinges on posterior summarization, which remains underexplored. To this end, we introduce the VC-measure (Variable Count and its rank variant) with a clustering-based threshold. This posterior summary is a simple, tuning-free plug-in that requires no sampling beyond the standard model fits used by existing methods, integrates with any BART variant, and avoids the instability of the median probability model and the computational cost of permutations. In a large-scale benchmark of 3,600 settings built on 100 nonlinear physics equations, it yields uniform $F_1$ gains for both general-purpose and sparsity-inducing priors; when paired with the Dirichlet Additive Regression Tree (DART), it overcomes pitfalls of the original summary and attains the best overall balance of recall, precision, and efficiency. Practical guidance on aligning summaries and downstream goals is discussed.
\end{abstract}


\section{Introduction}
Variable selection, or feature selection in machine learning, is the process of distilling the most relevant variables to include in a model. Owing to its benefits of improved interpretability and reduced overfitting, variable selection has attracted, and will continue to attract, significant attention in the statistics and machine learning communities. In its simplest form, variable selection is carried out in the context of linear regression, where the response variable or the outcome is assumed to be linearly associated with the features or predictors \citep{Bhattacharya02102015, horseshoe, SCAD, George01091993, spike_slab, LASSO, Zou2005}. However, such an assumption can be quite problematic when the features influence the outcome in a nonlinear manner, whose functional form is unknown \textit{a priori} \citep{Turlach2004}. Instead of first selecting a parametric model to filter out features, one can adopt a model-free point of view where we first identify the important variables and then construct a model, also known as \textit{nonparametric} or \textit{model-free} variable selection. 

Nonparametric variable selection has been studied through a variety of modeling frameworks, including basis expansions with coefficient shrinkage \citep{COSSO, interaction_ref_1, Ravikumar_2009}, derivative-based regularization \citep{RODEO}, the model-free knockoff framework for false discovery rate control \citep{knockoff}, sparsity-inducing Gaussian process priors \citep{GPVS}, and tree-based methods, a widely used class of nonparametric regression techniques in statistics and machine learning that is the focus of this paper. Tree-based methods have been developed from both the frequentist and Bayesian perspectives, often achieving state-of-the-art predictive performance while retaining interpretability through variable importance measures. Frequentist examples include random forests \citep{random_forest}, AdaBoost \citep{adaboost}, stochastic gradient boosting \citep{boosting}, and reinforcement learning trees \citep{Zhu02102015}, whereas Bayesian tree ensembles are built on Bayesian additive regression trees (BART) \citep{chipman2010} and their rich variants \citep{DART, ABC_forest}. Similar to most variable selection methods, variable selection for tree-based methods depends on two key components: a measure of \emph{variable importance} and a thresholding rule for deciding which variables to retain. Unlike the coefficient magnitudes in parametric models, however, variable importance in tree ensembles is less straightforward to define. For example, in random forests, the commonly used Gini importance \citep{gini} and permutation importance \citep{random_forest} can be unreliable when features vary in scale or category count \citep{Strobl2007}.

Compared to their frequentist counterparts, Bayesian tree-based methods retain competitive predictive performance and provide posterior samples, enabling variable selection based on probabilistic considerations. Among them, BART has emerged as a central framework on which most variable selection methods build. These methods fall into three broad categories: \emph{permutation-based approaches} \citep{bleich2014, BartMixVs} assess variable importance, typically via the variable inclusion proportion (VIP), and set selection thresholds from null distributions obtained by permuting the response; \emph{sparsity-inducing priors} \citep{DART, ABC_forest} modify the BART prior to encourage parsimonious ensembles, selecting variables using marginal posterior variable inclusion probabilities (MPVIP) and the median probability model (MPM) threshold \citep{MPM}; \emph{clustering-based approaches}~\citep{pan_sr} avoid tuning or permutation by separating relevant and irrelevant predictors through unsupervised clustering, often on rank-based importance measures, to achieve high recall. While each class offers distinct advantages, they differ markedly in accuracy, stability, and computational cost, and no single posterior summarization method dominates across settings.

In this paper, we modularize variable selection with Bayesian tree ensembles into two components: the tree prior and the posterior summary, and focus on the latter. While substantial effort has gone into designing new tree priors, the posterior summary has received disproportionately less attention. We argue that posterior summaries play a key role in nonparametric variable selection, a perspective reminiscent of the posterior-summarization view of parametric linear selection in \citet{Hahn2015}, but with substantial additional complexity in nonparametric tree ensemble methods. We introduce a new variable importance measure, the \emph{Variable Count} (VC), and its rank-based variant, \emph{VC Rank}, collectively termed the \emph{VC-measure}. The VC-measure avoids several pitfalls of existing posterior summaries. When paired with a clustering-based thresholding strategy, it requires no additional posterior sampling beyond standard model fitting, integrates seamlessly with any BART variant, and eliminates the computational burden of permutations. We evaluate the VC-measure against seven existing methods on the \emph{extended Feynman Symbolic Regression Database} \citep{pan_sr}, a benchmark of 100 nonlinear physics equations from the \emph{Feynman Lectures on Physics} \citep{Feynman_Lectures}, across 3,600 combinations of equation, sample size, and signal-to-noise ratio (SNR), each with 10 seeds, totaling 36,000 model fits per method. The VC-measure uniformly improves $F_1$ scores for both general-purpose and sparsity-inducing tree priors, and boosts the performance of DART \citep{DART} to achieve the best overall balance of recall, precision, and efficiency, all via a simple plug-in of the new posterior summary. Beyond the proposed method, the benchmark provides a systematic comparison of the accuracy, stability, and computational cost of existing methods and informs the choice of posterior summary for a given application. The proposed VC-measure is available as an open-source R package at
\href{https://github.com/mattsheng/BayesVC}{github.com/mattsheng/BayesVC}.

The remainder of the paper is organized as follows. Section~\ref{sec:nvs} defines the problem of nonparametric variable selection and describes the two classes of BART priors considered. Section~\ref{sec:posterior.summary} defines the existing posterior summaries we compare against, grouped by thresholding strategy: permutation-based, regularization-based, and clustering-based. Section~\ref{sec:cluster_vc} introduces the proposed VC-measure. Section~\ref{sec:sim} reports the benchmark results, and Section~\ref{sec:dis} concludes with recommendations for different application scenarios. Additional analyses are reported in Supplementary Material.

\section{Bayesian Tree Ensemble Methods for Nonparametric Variable Selection}
\label{sec:nvs}
Let $\by = (y_1, \ldots, y_n)^T \in \mathbb{R}^n$ denote a continuous response vector, and let $\bX = (\bx_1^T, \ldots, \bx_p^T) \in \mathbb{R}^{n \times p}$ denote the predictor (feature) matrix, where $\bx_j = (x_{1,j}, \ldots, x_{n,j})^T \in \mathbb{R}^n$ denotes the $j$th predictor, for $j = 1,\ldots,p$. We consider the canonical nonparametric regression problem, where the continuous response vector $\by$ is associated with the $p$ predictors $\bX$ through an unknown data-generating function $f_0(\cdot)$:
\begin{equation} \label{eq:np}
    \by = f_0(\bX) + \bm\varepsilon, \qquad \bm\varepsilon \sim \mathcal{N}(\bm{0}, \sigma^2\bm{I}_n),
\end{equation}
such that $\sigma^2 \in \mathbb{R}^+$ is the noise variance. In the purview of nonparametric variable selection, it is assumed that only a small subset $\mathcal{S}_0 \subseteq \{1, \ldots, p\}$ of $p_0 = |\mathcal{S}_0|$ predictors exerts influence on $\by$. That is, the nonparametric regression problem in \eqref{eq:np} can be further simplified to $\by = f_0(\Xrel) + \bm\varepsilon$, where $\Xrel = (\bx_j)_{j \in \mathcal{S}_0} \in \mathbb{R}^{n \times p_0}$ denotes the relevant predictor matrix. Given the dataset $(\by, \bX)$, one wishes to recover the active set $\mathcal{S}_0$. For convenience, we define $\mathcal{S}_1 = [p] \setminus \mathcal{S}_0$ and $\Xirr$ denotes the irrelevant predictor matrix.

The proposed posterior summary applies to two classes of Bayesian tree priors: \textit{general BART} (a general-purpose nonparametric regression prior) and \textit{sparsity-inducing BART} (variants of general BART that explicitly encourage sparsity to achieve parsimonious models). We describe both classes next and show in Section~\ref{sec:sim} that the VC-measure improves both, often uniformly across the considered settings.

\subsection{General BART} 
\label{sec:BART}
BART is a Bayesian ensemble model for nonparametric function estimation and inference, motivated by ensemble methods in general and boosting algorithms in particular. For the regression problem in~\eqref{eq:np}, BART models $f_0$ by a sum of $T$ regression trees,
\begin{equation} \label{eq:sum-of-trees}
    f_0(\bX) \approx \sum_{t=1}^T g_t(\bX; \mathcal{T}_t, \bm\mu_t),
\end{equation}
where the binary regression tree $\mathcal{T}_t$ consists of a set of $b_t$ terminal nodes with parameters $\bm\mu_t = (\mu_{t, 1},\ldots,\mu_{t, b_t})$ and a set of decision rules, each a binary split of the predictor space of the form $\{x_j \leq c\}$ vs $\{x_j > c\}$ for some $j = 1,\ldots,p$. Owing to the partition nature of $\mathcal{T}_t$, each sample $\bx = (x_1,\ldots,x_p)$ is associated with exactly one terminal node of $\mathcal{T}_t$, and $g_t(\bx; \mathcal{T}_t, \bm\mu_t) = \mu_{t,\ell}$ if $\bx$ falls into the $\ell$th terminal node. The sum-of-trees model \eqref{eq:sum-of-trees} is then the sum of $T$ terminal node values, each dictated by a binary tree.

The number of trees $T$ is a prespecified hyperparameter that trades off representation flexibility against competition among predictors. A large $T$ improves predictive performance but allows irrelevant features to enter the model without worsening the overall fit, making variable selection unnecessarily more challenging~\citep{bleich2014}. A small $T$ forces predictors to compete for the few available splits, which is better suited for variable selection, but when $T$ is too small, BART lacks the representation power to estimate a complicated regression function $f_0$ and can be trapped in local modes~\citep{bayesian_cart}. \cite{chipman2010} suggest $T=200$ for prediction and inference, whereas $T=20$ is a common choice for variable selection \citep{bleich2014, ABC_forest, BartMixVs}.

To prevent any single tree from dominating the additive structure of the sum-of-trees model \eqref{eq:sum-of-trees}, a regularization prior is imposed over all model parameters, namely, the $T$ independent binary trees $\{\mathcal{T}_t\}_{t=1}^T$, the set of terminal parameters $\{\bm\mu_t\}_{t=1}^T$, and the error variance $\sigma^2$. The prior gives strong preferences to small and simple trees, and assumes the tree components $(\mathcal{T}_t, \bm\mu_t)$ are independent of each other and of $\sigma^2$. This prior independence enables an efficient Metropolis-within-Gibbs sampler based on Bayesian backfitting~\citep{bayesian_backfitting}, which entails $T$ successive draws of $(\mathcal{T}_t, \bm\mu_t)$ followed by a draw of $\sigma^2$. Specifically, let $\bm{r}_t = \bm{y} - \sum_{m \neq t} g(\bm{X}; \mathcal{T}_m, \bm\mu_m)$ denote the $n$-dimensional partial residual vector based on the fit that excludes the $t$th tree. The tree $\mathcal{T}_t$ is drawn from $\mathcal{T}_t \mid \bm{r}_t, \sigma^2$ using the Metropolis-Hastings (MH) algorithm introduced in \cite{bayesian_cart}, which proposes a new tree $\mathcal{T}^*$ from the current tree $\mathcal{T}$ using one of four moves: BIRTH (growing a terminal node), DEATH (pruning a pair of terminal nodes), CHANGE (changing a nonterminal rule), and SWAP (swapping a rule between parent and child). Then, $\bm\mu_t$ is drawn from its Gaussian full conditional given $(\mathcal{T}_t, \bm{r}_t, \sigma^2)$. Once all $T$ pairs of $(\mathcal{T}_t, \bm\mu_t)$ are drawn, $\sigma^2$ is sampled from its inverse-Gamma full conditional. The full prior specification and posterior sampling algorithm are given in \cite{chipman2010}.

\subsection{Sparsity-inducing Regularization Bayesian Trees}
\label{sec:reg}
The general BART is a flexible machinery for nonparametric regression. For variable selection, as discussed in Section~\ref{sec:BART}, using a small number of trees increases competition among predictors and can make general BART more suitable for the task. Alternatively, one can modify the BART prior by introducing regularization to encourage sparsity in the tree ensembles. We consider two such methods, which differ in how regularization is incorporated into the prior. DART replaces the uniform variable splitting probability with a sparse prior, while \abc wrap a sparse prior directly around the variables available to the ensemble. In essence, DART's strategy mimics a soft thresholding, and that of \abc resembles a hard thresholding.

\subsubsection{Dirichlet Additive Regression Tree}
\label{sec:reg_DART}

The tree prior $p(\mathcal{T})$ in BART follows from \cite{bayesian_cart} and is specified through a tree-generating stochastic process. Starting from a root node at depth $d=0$, a node at depth $d$ is split with probability
\begin{equation} \label{eq:p_split}
    p_{\text{split}}(d) = \frac{\gamma}{(1+d)^\beta}, \qquad \gamma \in (0,1), \qquad \beta \in [0, \infty),
\end{equation}
in which case it is assigned a splitting rule of the form $\{\bx_j \leq c\}$ vs $\{\bx_j > c\}$ for some $j = 1,\ldots,p$ and some $c$ among the observed values of $\bx_j = (x_{1,j}, \ldots, x_{n,j})$, giving two child nodes of depth $d+1$; otherwise, the node is terminated with a terminal value $\mu$. A predictor whose observed values are exhausted, which is common for binary or categorical predictors, is no longer eligible at the nodes below, and a node is terminated when no nontrivial splitting rule remains. With the default choice of $\gamma = 0.95$ and $\beta = 2$ in \eqref{eq:p_split}, the prior puts most probability on tree depths of 2 or 3, limiting each tree in the ensemble to be a weak learner. Even so, \cite{DART} note that such regularization is insufficient when $p$ is large since the splitting rule assignment is not regularized: the predictor used to construct a splitting rule is chosen according to the probability vector $\bm{s} = (s_1,\ldots,s_p)$, which is set to $\bm{s} = (1/p, \ldots, 1/p)$ by default in BART. In high-dimensional settings, they show that this uniform specification leads to a preference for highly nonsparse models. Instead, \cite{DART} propose Dirichlet Additive Regression Tree (DART) by replacing the uniform prior for $\bm{s}$ with a sparsity-inducing Dirichlet prior:
\begin{equation}\label{eq:s_prior}
    \bm{s} \sim \mathcal{D}\left(\frac{\alpha}{p}, \ldots, \frac{\alpha}{p}\right),
\end{equation}
where $\alpha \in \mathbb{R}^+$ is a hyperparameter controlling the sparsity of the model. Under a mild assumption, \cite{DART} show that DART favors a tree ensemble much sparser than BART. In a fully Bayesian specification, one can place a hyperprior on $\alpha$ of the form $\alpha/(\alpha + \rho) \sim \text{Beta}(a,b)$, where $(a, b, \rho)$ are hyperparameters, and the default choice of $(a,b,\rho) = (0.5, 1, p)$ provides a suitable sparsity level for most cases.

When the splitting rules $\{\bx_j \leq c\}$ vs $\{\bx_j > c\}$ of all predictors are not exhausted, e.g., when all predictors are continuous, the Dirichlet prior in \eqref{eq:s_prior} gives a simple conjugate Gibbs update $\bm{s} \sim \mathcal{D}(\alpha/p + c_1, \ldots, \alpha/p + c_p)$, where $c_j$ is the number of $\bx_j$-splitss in the ensemble; otherwise, this update can serve as the proposal distribution of an independent Metropolis-Hastings step for $\bm{s}$.

\subsubsection{\abc}
\label{sec:reg_ABC}

Similar in spirit to DART, \cite{ABC_forest} propose \abc by explicitly adding regularization to the BART prior. Instead of regularizing the variable splitting probability $\bm{s}$, they place the spike-and-forest prior \citep{BART_post_concentration} directly on the variable index set $\mathcal{S} \subseteq \{1,\ldots,p\}$, thereby limiting which predictors are available for tree construction: $\mathcal{S} \sim \pi(\mathcal{S})$ and, given $\mathcal{S}$, the tree components $(\mathcal{T}_1, \bm\mu_1), \ldots, (\mathcal{T}_T, \bm\mu_T)$ and $\sigma^2$ follow the BART prior restricted to $\bX_{\mathcal{S}}$. A typical and practical choice of $\pi(\mathcal{S})$ is the beta-binomial prior $\theta_0 \sim \text{Beta}(a_0, b_0)$, $\gamma_j \mid \theta_0 \sim \text{Bernoulli}(\theta_0)$ independently for $j = 1, \ldots, p$, and $\mathcal{S} = \{j : \gamma_j = 1\}$, with default $(a_0, b_0) = (1, 1)$. However, the spike-and-forest prior renders the marginal likelihood $p(\by \mid \mathcal{S})$ intractable. Thus, \cite{ABC_forest} propose an approximate Bayesian computation (ABC) algorithm that obviates its direct computation, together with a data-splitting strategy to improve the acceptance rate of the ABC sampler. Specifically, at each ABC iteration $\ell = 1,\ldots,\Labc$, a subset $\mathcal{S}_\ell$ of predictors is sampled from the prior $\pi(\mathcal{S})$, and the observations are randomly split into two partitions $\mathcal{I}_\ell$ and $\mathcal{I}_\ell^c$ (of equal size by default). A BART model is then trained on $(\by_{\mathcal{I}_\ell}, \bX_{\mathcal{I}_\ell, \mathcal{S}_\ell})$, and the ABC sample is accepted if $\varepsilon_\ell = \norm{\by_{\mathcal{I}_\ell^c} - \tilde{\by}_{\mathcal{I}_\ell^c}}_2$, the $\ell_2$-discrepancy between the held-out responses and pseudo-responses $\tilde{\by}_{\mathcal{I}_\ell^c}$ drawn from the posterior predictive of the fitted model, is less than a prespecified threshold $\varepsilon_0$, and is rejected otherwise. In practice, $\varepsilon_0$ is difficult to determine. \cite{ABC_forest} propose to keep only a fraction of ABC samples (top 10\% in $\varepsilon_\ell$ by default) for posterior inference.

\section{Posterior Summary: Variable Importance Measure for Variable Selection}\label{sec:posterior.summary}

Posterior summaries translate posterior samples into a sparse representation, a necessary step under any Bayesian tree prior. Posterior summarization involves two key components: a variable importance measure and a threshold for variable selection. Common variable importance measures in the literature include VIP~\citep{bleich2014} and MPVIP~\citep{DART,ABC_forest}, as well as more recent approaches such as VIP Rank~\citep{pan_sr} and Metropolis importance (MI)~\citep{BartMixVs}. Threshold selection is likewise an integral part of the summarization procedure, with common choices including permutation-based thresholds~\citep{bleich2014}, fixed cutoff of 0.5 for MPM, and tuning-free clustering-based thresholds~\citep{pan_sr}.

Although a variable importance measure can, in principle, be paired with any of the tree priors discussed in Section~\ref{sec:nvs}, there is often a natural synergy between the chosen tree prior and the way variable importance is computed, so existing measures are generally not transferable across BART priors. We describe the existing methods that serve as competitors in Section~\ref{sec:sim}, each defined by its combination of tree prior and posterior summary, and group them according to their shared thresholding strategy. In contrast, the variable importance measure proposed in Section~\ref{sec:cluster_vc} is designed to be versatile, offering improvements for both general-purpose and sparsity-inducing BART.

\subsection{Permutation-based Variable Selection for General BART}
\label{sec:perm}

Permutation-based methods determine the selection threshold for a given variable importance measure from its null distribution under permuted responses. We first describe the framework of \cite{bleich2014} which was built on VIP, and then its extension to the Metropolis importance of \cite{BartMixVs}.

\subsubsection{BART VIP}
\label{sec:perm_vip}

The tree structures $\{\mathcal{T}_t\}_{t=1}^T$ in the posterior samples from the Markov chain Monte Carlo (MCMC) algorithm are essential to all Bayesian tree ensemble variable selection methods, as they encode which predictors are split on and where. Intuitively, if the response $\by$ changes rapidly with respect to $\bx_j$, the ensemble needs many splits on $\bx_j$ to capture this behavior, so $\bx_j$ is often viewed as a more important driver of $\by$ than $\bx_i$ if the ensemble splits on $\bx_j$ more frequently than on $\bx_i$. Counting the number of times $\bx_j$ is being split on (or $\bx_j$-splits hereinafter) among all splitting rules in the ensemble, normalizing by the total number of splitting rules, and averaging across posterior samples yields the \emph{variable inclusion proportion (VIP)} $q_j$ for predictor $\bx_j$.

\begin{definition}[BART variable inclusion proportion]\label{def:vip}
    Let $K$ denote the number of posterior samples obtained from a BART model post burn-in period. For each $j = 1,\ldots,p$ and $k = 1,\ldots,K$, let the variable count $c_{j,k}$ be the number of $\bx_j$-splits in the $k$th posterior sample and let $c_{\cdot,k} = \sum_{j=1}^p c_{j,k}$ denote the total number of splitting rules in the $k$th posterior sample. The \emph{variable inclusion proportion} (VIP) for predictor $\bx_j$ is defined as
    \begin{equation}\label{eq:vip}
        q_j = \frac{1}{K}\sum_{k=1}^K \frac{c_{j,k}}{c_{\cdot,k}}, \qquad j = 1,\ldots, p.
    \end{equation}
\end{definition}

\cite{chipman2010} suggest that BART VIPs $\bm q = (q_1,\ldots,q_p)$ can be used to rank predictors in terms of relative importance, but a complete and principled selection criterion is not given. \cite{bleich2014} notice that BART VIPs vary considerably over multiple independent BART fits on the same dataset, so a single fit cannot be relied upon to determine an appropriate selection threshold for $q_j$. It is then customary to run $\Lrep > 1$ ($\Lrep = 10$ by default) independent BART fits and average the posterior summary, BART VIP in this case, for a robust assessment of each variable's importance~\citep{bleich2014,BartMixVs}.

The key to an appropriate threshold for $q_j$ lies in the behavior of $q_j$ when $\bx_j$ is not associated with $\by$, i.e., the null distribution of $q_j$. \cite{bleich2014} propose to estimate the null distributions by permuting the response vector $\by$ so that the permuted response is no longer associated with any predictor. Since the number of possible permutations, $n!$, is prohibitively large, $\Lperm$ permuted responses $\{\by_\ell^\star\}_{\ell=1}^\Lperm$ are created, where $\Lperm$ is a prespecified hyperparameter typically set to 50--100. A BART model is then fitted independently to each null dataset $(\by_\ell^\star, \bX)$. For each predictor $\bx_j$, the resulting null VIPs $\bq_j^\star = (q_{j,1}^\star,\ldots,q_{j,\Lperm}^\star)$ is a sample from the null distribution of $q_j$, i.e., the behavior of $q_j$ as if $\bx_j$ were not associated with $\by$. A permutation test then determines a selection threshold: the null hypothesis is that $q_j$ comes from the null distribution $\bq_j^\star$, and the alternative is that $q_j$ is significantly larger than $\bq_j^\star$, i.e., $\bx_j$ is associated with $\by$. As noted above, $q_j$ is estimated as the average BART VIP of $\bx_j$ over $\Lrep$ independent fits on the original dataset $(\by, \bX)$.

\cite{bleich2014} introduce three strategies of varying selection stringency for determining the thresholds for $q_j$'s. The least stringent one, \Local, requires $q_j$ to be extreme with respect to its own null distribution. That is, $\bx_j$ is selected if $q_j > \mathcal{Q}_{j, 1-\alpha}^\star$, the $1-\alpha$ sample quantile of $\bq_j^\star$. This permutation test is conducted independently for all $p$ predictors and therefore offers no control of multiple testing errors such as the family-wise error rate (FWER) or the false discovery rate (FDR). The most stringent one, \Gmax, calculates the threshold from the largest null BART VIP across all predictors: let $q_{\max, \ell}^\star = \max_{1 \leq j \leq p} q_{j,\ell}^\star$ for $\ell = 1, \ldots, \Lperm$, and let $\mathcal{Q}_{\max, 1-\alpha}^\star$ denote the $1-\alpha$ sample quantile of $\{q_{\max,1}^\star, \ldots, q_{\max, \Lperm}^\star\}$; then $\bx_j$ is selected if $q_j > \mathcal{Q}_{\max, 1-\alpha}^\star$. The third strategy, \Gse, strikes a balance between the former two. Let $m_j$ and $s_j^2$ denote the sample mean and variance of the null VIPs $\bq_j^\star$, respectively. Define
\begin{equation}
    C^\star = \inf_{C \in \mathbb{R}^+} \left\{\forall\, j, \frac{1}{\Lperm}\sum_{\ell=1}^\Lperm \mathbbm{1}(q_{j,\ell}^\star \leq m_j + C \cdot s_j) > 1-\alpha \right\},
\end{equation}
the smallest global multiplier that gives simultaneous $1-\alpha$ coverage across the permutation distributions of all predictors. A predictor $\bx_j$ is then selected if $q_j > m_j + C^\star \cdot s_j$. Unlike \Local, whose threshold depends only on the predictor's own null distribution, \Gse incorporates information from the null distributions of other predictors through $C^\star$, but does so in a less stringent way than \Gmax. We show in Section~\ref{sec:comparison} that \Gse indeed has the highest $F_1$ score among the three criteria proposed by \cite{bleich2014} in all but the noiseless settings with $n \geq 1500$.

These selection rules are, however, computationally burdensome since they require $\Lrep + \Lperm$ independent BART fits. In our benchmark (Section~\ref{sec:runtime}), the three criteria are the second slowest methods, ahead of only \MI. Nonetheless, one can have three different sets of selected variables at once.

\subsubsection{BART Metropolis Importance}
\label{sec:perm_MI}

The BART Metropolis importance (\MI) is a variable importance measure proposed by \cite{BartMixVs} to address BART VIP's bias against low-cardinality predictors (e.g., binary and categorical predictors), but it is also a suitable measure for datasets with only continuous predictors. Instead of building on the variable count $c_j$, \cite{BartMixVs} construct the measure from $\pi_{\text{BIRTH}}(\eta) = \min\{1, r(\eta)\}$, the Metropolis ratio for accepting a BIRTH proposal at a terminal node $\eta$ of tree $\mathcal{T}$, where $r(\eta) = P(\mathcal{T} \mid \mathcal{T}^*)P(\mathcal{T}^* \mid \bm{r}, \sigma^2) / \{P(\mathcal{T}^* \mid \mathcal{T})P(\mathcal{T} \mid \bm{r}, \sigma^2)\}$. The intuition is that a predictor $\bx_j$ may be more important if a terminal node $\eta$ is more likely to become a splitting rule on $\bx_j$ than on the other predictors. Since $\pi_{\text{BIRTH}}(\eta)$ cannot be used directly as a variable importance measure, \cite{BartMixVs} propose the following normalization.

\begin{definition}[BART Metropolis importance]
   Let $\phi_{t,k}$ denote the set of interior nodes of the $t$th tree $\mathcal{T}_t$ in the $k$th posterior sample. Define the average Metropolis acceptance ratio per $\bx_j$-splitsting rule in the $k$th posterior sample as follows:
    \begin{equation} \label{eq:normalized_MR}
        \tilde{u}_{j,k} = \frac{\sum_{t=1}^T\sum_{\eta \in \phi_{t,k}} \mathbbm{1}(j_\eta = j)\pi_{\text{BIRTH}}(\eta)}{\sum_{t=1}^T \sum_{\eta \in \phi_{t,k}} \mathbbm{1}(j_\eta = j)}, \qquad\forall\, j = 1,\ldots,p, \quad\forall\, k = 1,\ldots,K,
    \end{equation}
    where $j_\eta$ denotes the variable index of the split variable at the interior node $\eta$. That is, $\mathbbm{1}(j_\eta = j)$ indicates whether $\eta$ is a $\bx_j$-splits or not. Set $\tilde{u}_{j,k} = 0$ whenever the denominator of \eqref{eq:normalized_MR} is 0. The BART Metropolis importance for predictor $\bx_j$ is defined as 
    \begin{equation}
        v_j^{\text{MI}} = \frac{1}{K}\sum_{k=1}^K \frac{\tilde{u}_{j,k}}{\sum_{i=1}^p \tilde{u}_{i,k}}, \qquad\forall\, j = 1,\ldots, p,
    \end{equation}
    the normalized $\tilde{u}_{j,k}$ averaged across $K$ posterior samples.
    \end{definition}

To determine thresholds for $v_j^{\text{MI}}$, \cite{BartMixVs} adapt the permutation framework of \cite{bleich2014} by replacing $q_j$ with the median of $v_j^{\text{MI}}$ over the $\Lrep$ fits, requiring $\Lrep + \Lperm$ independent BART fits. Currently, the R package \texttt{BartMixVs} only supports the ``Local'' selection criterion, and we refer to this method as \MI. Under mixed-type predictors, \cite{BartMixVs} find that \MI achieves no worse true positive rate (TPR) than \Local in most settings but at a slightly lower precision, leading to a marginally lower $F_1$ score. In our simulations with continuous predictors (Section~\ref{sec:FP.TP}), \MI has comparable TPR to \Local in noisy settings, but a consistently higher false positive rate (FPR). Moreover, it selects no variables in large-$n$, low-noise settings, resulting in markedly lower $F_1$ scores than \Local in most of the simulation settings considered.

\subsection{Median Probability Model for Sparsity-inducing BART}
\label{sec:reg_vs}
Regularization Bayesian tree priors in Section~\ref{sec:reg} have built-in sparsity in the tree ensembles. Despite differing in how regularization is incorporated into the BART prior, they use the same variable importance measure and posterior summary. 
\subsubsection{Dirichlet Additive Regression Tree (DART)}
\label{sec:reg_DART_vs}

By treating each posterior draw of the tree ensemble as an independent model, DART uses MPVIP
\begin{equation}
    \pi_j = P(\bx_j \text{ is in the model} \mid \by), \qquad j = 1,\ldots,p,
\end{equation}
and selects predictors whose MPVIP is at least 0.5, yielding the median probability model \citep{MPM}. In practice, $\pi_j$ is estimated as follows:
\begin{equation}\label{eq:mpvip_hat}
    \hat{\pi}_j = \frac{1}{K}\sum_{k=1}^K \mathbbm{1}\left(\bx_j \in \{\mathcal{T}_{t,k}\}_{t=1}^T\right), \qquad j = 1,\ldots,p,
\end{equation}
where $K$ is the number of posterior draws and $\mathbbm{1}\left(\bx_j \in \{\mathcal{T}_{t,k}\}_{t=1}^T\right)$ indicates whether $\bx_j$ has been split on \emph{at least once} among all $T$ trees in the $k$th posterior draw.

There is no conceptual obstacle to coupling BART with MPVIP. However, without proper regularization, MPVIP is not a suitable variable importance measure since the requirement of \emph{only} one occurrence of $\bx_j$ in a posterior draw can substantially inflate $\hat{\pi}_j$. In unreported results, BART with the MPVIP importance measure and the MPM selection criterion consistently yields the worst performance in all simulations considered.

The explicit MPM threshold of $\pi_j \geq 0.5$ means no permuted fit is needed for threshold estimation, making it the most computationally efficient method at $\Lrep = 1$ among those considered in this article. Empirically, we find that \dart has inadequate regularization when the error variance $\sigma^2$ is small, possibly due to posterior overconcentration. In particular, a subset of irrelevant variables $\Xirr$ is found to have low split frequencies (typically 1 to 3 splits within an ensemble) persistently over 50\% of the posterior draws, thereby inflating their MPVIPs to over 0.5 and leading to false discoveries. See Sections~\ref{sec:comparison} and \ref{sec:FP.TP} for more details.

\subsubsection{\abc}
\label{sec:reg_ABC_vs}

Variable selection for \abc is also achieved by examining the MPVIPs and selecting those with $\hat{\pi}_j \geq 0.5$. Where they differ is that $\hat{\pi}_j$ in \abc is calculated using only the top 10\% ABC samples retained (Section~\ref{sec:reg_ABC}). Similar to the permutation-based methods, \abc require multiple independent BART fits ($\Labc$ in total), but each fit requires a much smaller burn-in period (200 by default) and only a \emph{single} posterior draw. With complicated datasets, a larger burn-in period may be required to ensure better mixing and thereby a better ABC proposal. In general, we find that \abc have comparable $F_1$ scores to the BART VIP methods described in Section~\ref{sec:perm_vip}, but they tend to miss relevant predictors, a phenomenon independently observed in \cite{BartMixVs}.

\subsection{Clustering-based Variable Selection}
\label{sec:cluster_vip_rank}
\vipRank~\citep{pan_sr} is a permutation- and regularization-free variable selection method for general BART with a focus on ensuring that all relevant predictors $\Xrel$ are selected. Although designed as a pre-screening filter for high-dimensional symbolic regression problems, it is also suitable for other pre-screening tasks where the recovery of $\Xrel$ outweighs the exclusion of irrelevant predictors $\Xirr$. Inspired by BART VIP methods where $q_j$'s gauge the relative importance of each feature, \cite{pan_sr} advocate a new variable importance measure, VIP Rank, which uses the ranking of $q_j$ rather than its raw value. In contrast to $q_j$, ranks are not on an arbitrary scale, so the permutation technique is not needed to determine an appropriate threshold. Specifically, let $R(q_j)$ denote the ranking of $q_j$ among $(q_1,\ldots,q_p)$ in descending order. Since relevant features $\Xrel$ tend to have larger VIP values than irrelevant features $\Xirr$, which are only split on sporadically or by chance in the BART ensemble~\citep{bleich2014, chipman2010}, the rankings of $\Xrel$ occupy the top-ranking positions $\{1,\ldots,p_0\}$, whereas those of $\Xirr$ appear in the lower-ranking positions $\{p_0 + 1, \ldots, p\}$. Selecting $\bx_j$ whenever $R(q_j) \leq p_0$ is, however, impractical as the sparsity $p_0$ is generally unknown, and the single-fit ranking vector $\big(R(q_1),\ldots,R(q_p)\big)$ is just a permutation of the variable indices $(1,\ldots,p)$ that provides no insight into $p_0$. Instead, \cite{pan_sr} find that VIP rankings over multiple independent BART fits provide differential characterization of relevant rankings versus irrelevant rankings, leading to an intuitive variable selection strategy.

\begin{definition}[\vipRank]\label{def:vip_rank}
    Let $q_{j,\ell}$ denote the BART VIP for predictor $\bx_j$ in the $\ell$th fit. Within each model fit $\ell$, denote $R(q_{j,\ell})$ the rank of $q_{j,\ell}$ among $(q_{1, \ell}, \ldots, q_{p, \ell}) \in \mathbb{R}^p$ in descending order. Define the \vipRank for predictor $\bx_j$ as the average of $\big(R(q_{j,1}), \ldots, R(q_{j,\Lrep})\big) \in \mathbb{R}^{\Lrep}$ over $\Lrep$ independent fits:
    \begin{equation}\label{eq:vip_rank}
        \overline{R}^q_j = \frac{1}{\Lrep}\sum_{\ell=1}^\Lrep R(q_{j,\ell}), \qquad j = 1,\ldots,p.
    \end{equation}
\end{definition}

Since each of the $\Lrep$ BART models is fitted independently, the rankings of the same predictor $\bx_j$ over different runs, $\big(R(q_{j,1}), \ldots, R(q_{j,\Lrep})\big)$, can be seen as independent samples, and empirical evidence in \cite{pan_sr} suggests that the relevant versus irrelevant per-run rankings are random variables with different supports. In particular,
\begin{equation}\label{eq:vip_rank_assumption}
    R(q_{j, 1}), \ldots, R(q_{j,\Lrep}) \overset{\text{iid}}{\sim} \begin{cases}
        \text{Unif}(\{1, \ldots, p_0\}), &\text{if } j \in \mathcal{S}_0,\\
        \text{Unif}(\{p_0+1, \ldots, p\}), &\text{otherwise,}
    \end{cases}
\end{equation}
so that $\overline{R}^q_j$ concentrates around $(1+p_0)/2$ if $j \in \mathcal{S}_0$ and around $(p_0+1+p)/2$ otherwise, where $(1+p_0)/2 \ll (p_0+1+p)/2$ in sparse settings. Although these values are unknown, they are well-separated and the separation directly corresponds to predictor relevance. To illustrate, consider a regression problem with $p=204$ features, of which $p_0=4$ are relevant; then $\Xrel$ cluster around $(1+p_0)/2 = 2.5$, while $\Xirr$ form a cluster with mean $(p_0+1+p)/2 = 104.5$.

Owing to this separation, \cite{pan_sr} propose to frame variable selection as a clustering problem over the VIP ranks: a clustering algorithm is applied to the VIP Rank vector $\big(\overline{R}^q_1, \ldots, \overline{R}^q_p\big)$ to identify two clusters, and a predictor $\bx_j$ is selected if $\overline{R}^q_j$ belongs to the cluster with the smaller average VIP Rank (denoted the low-mean cluster), and is discarded otherwise. This seemingly trivial clustering problem, however, presents a unique challenge due to the intrinsic sparsity assumption, $p_0 \ll p$, which naturally creates a \emph{class imbalance} problem. In particular, \cite{pan_sr} find that centroid-based and density-based clustering methods struggle in this scenario since the larger cluster of irrelevant predictors ($p-p_0$ instances) can dominate centroid positions and overshadow the density signals of the smaller cluster of relevant predictors ($p_0$ instances). In contrast, hierarchical agglomerative clustering (HAC) is much more robust to class imbalance since it starts with each point as its own cluster and merges them based purely on their pairwise distances. To this end, \cite{pan_sr} propose to perform HAC on the VIP Rank vector using Euclidean distance and average linkage, cut the resulting dendrogram to form exactly two clusters, and retain the low-mean cluster, i.e., the higher-ranking predictors, as the relevant set.

The only tuning parameter in \vipRank is the number of model replicates $\Lrep$, and the default choice is $\Lrep = 20$. Similar to the permutation-based methods in Section~\ref{sec:perm}, all BART replicates are independent and can be parallelized. \vipRank is more efficient than permutation-based methods owing to the fewer model replicates required (see Section~\ref{sec:runtime}) and has the highest TPR in all simulation settings by significant margins (see Section~\ref{sec:FP.TP}), making it ideal for tasks where the recovery of relevant predictors $\Xrel$ is critical. In noisy settings, however, some irrelevant predictors frequently occupy high-ranking positions (i.e., close to but larger than $p_0$), making \vipRank prone to false discoveries.

\section{Proposed Posterior Summarization: Variable Count and Its Rank}
\label{sec:cluster_vc}

\cite{pan_sr} point to a new avenue for BART-based variable selection by leveraging ranking statistics to transform variable importance measures on an arbitrary scale into interpretable and separable quantities. By evaluating relative standing rather than raw magnitude, this approach effectively distinguishes $\Xrel$ from $\Xirr$, resulting in a simple and computationally efficient algorithm that reliably recovers all relevant features. However, its strong TPR performance often comes at the expense of frequent false positives, a costly trade-off in general-purpose variable selection tasks.

In this work, we propose a new suite of variable importance measures, variable count (VC) and variable count rank (VC Rank), collectively referred to as \emph{VC-measure} hereinafter, along with a clustering-based thresholding procedure. The proposed posterior summary method has the following features:
\begin{itemize}
    \item Simple. The VC-measure operates entirely as a posterior summarization step, requiring no additional posterior samples beyond those already obtained from the MCMC run, and its clustering-based thresholding avoids the need for tuning-intensive procedures.
    \item Versatile. The VC-measure can be paired with any BART variant, including general BART and sparsity-inducing variants such as DART and ABC Bayesian forests, as a drop-in alternative to existing variable importance measures such as VIP, MPVIP, MI, and VIP Rank.
    \item Fast. Clustering-based thresholding removes the need for computationally intensive permutation-based thresholds; as shown in Section~\ref{sec:superior_of_VC}, when paired with BART, the proposed summary uniformly improves upon permutation-based BART.
    \item Accurate. Extensive numerical experiments in Section~\ref{sec:sim} show that, when paired with BART or DART, the VC-measure yields uniformly higher $F_1$ scores than other posterior summaries for these models across all simulation settings; when combined with DART, it achieves the highest $F_1$ scores across all settings with a more balanced trade-off between recall and precision, and it avoids the false positive inflation observed with \vipRank.
\end{itemize}

\begin{definition}[Variable Count (VC)]\label{def:vc}
    For each predictor $\bx_j$ and model fit $\ell = 1,\ldots, \Lrep$, let $c_{j, \ell} = \frac{1}{K}\sum_{k=1}^K c_{j,\ell,k}$ denote the average number of $\bx_j$-splits across all $K$ posterior draws in the $\ell$th model fit. Define the variable count (VC) vector for predictor $\bx_j$ as $\bm{c}_j = (c_{j,1}, \ldots, c_{j, \Lrep}) \in \mathbb{R}^{\Lrep}$. Stacking the VC vectors row-wise gives the VC matrix $\bm{C} = (\bm{c}_1^T, \ldots, \bm{c}_p^T)^T \in \mathbb{R}^{p \times \Lrep}$ for all predictors.
\end{definition}

\begin{definition}[Variable Count Rank (VC Rank)]\label{def:vc_rank}
    Within each model fit $\ell$, rank $(c_{1, \ell}, \ldots, c_{p, \ell}) \in \mathbb{R}^p$ in descending order and let $R(c_{j,\ell})$ denote the rank of predictor $\bx_j$ in fit $\ell$. Then, the VC Rank vector for predictor $\bx_j$ is defined as $\bm{R}_j^c = (R(c_{j,1}), \ldots, R(c_{j, \Lrep})) \in \{1,\ldots,p\}^\Lrep$, and stacking the VC Rank vectors row-wise gives the VC Rank matrix $\bm{R}^c = ((\bm{R}_1^c)^T, \ldots, (\bm{R}_p^c)^T)^T \in \{1,\ldots,p\}^{p \times \Lrep}$ for all predictors.
\end{definition}

We note that VC is a special case of VIP defined in \eqref{eq:vip}. To see this, we omit the model index $\ell$ for notational simplicity and denote the VC for predictor $\bx_j$ by $c_j = \frac{1}{K}\sum_{k=1}^K c_{j,k}$. Recall that the VIP for predictor $\bx_j$ is defined as $q_j = \frac{1}{K}\sum_{k=1}^K \frac{c_{j,k}}{c_{\cdot,k}}$, where $c_{\cdot,k} = \sum_{j=1}^p c_{j,k}$ is the total number of splitting rules in the $k$th posterior draw. Thus, VC can be regarded as VIP by setting the normalizing constant $c_{\cdot,k}$ to 1 for all predictors and all posterior draws. 

While VIP's normalization facilitates comparison across predictors within a given draw, it naturally imposes a fixed-sum constraint in that VIPs across $p$ features must sum to 1, which can obscure meaningful signal differences. For instance, when the model occasionally splits on irrelevant predictors $\Xirr$, a behavior not uncommon in BART \citep{bleich2014}, VIPs of relevant features $\Xrel$ are necessarily reduced. The relative scaling thus leads to diminished separation between signal and noise. Moreover, when $p$ is large, the normalization further flattens the VIP distribution, making it increasingly difficult to distinguish $\Xrel$ from $\Xirr$.

By focusing on the absolute scale rather than relative frequency, VC removes the fixed-budget constraint inherent in VIP: $\Xirr$ typically appear infrequently and sporadically in splitting rules, so their VC values remain low and do not dilute the signal of $\Xrel$, whose importance remains stable and distinguishable. Although VIP Rank also avoids the pitfalls of normalization, it considers only the relative ordering within a constrained, normalized system and ignores how often variables are selected. An empirical comparison of VC- and VIP-measure on identical DART posterior samples is given in Section C of Supplementary Material.

To select variables, we adapt the clustering framework described in Section~\ref{sec:cluster_vip_rank}. While one could apply HAC directly to the $p \times \Lrep$ matrices $\bm{C}$ and $\bm{R}^c$, doing so is ineffective in practice since each column, a single model fit, would be treated as an independent clustering feature, and these columns are noisy and lack discriminatory power; clustering in the full $p \times \Lrep$ space thus introduces high-dimensional noise that obscures the separation between $\Xrel$ and $\Xirr$.

To address this, we reduce each row of $\bm{C}$ and $\bm{R}^c$ by extracting low-dimensional summary statistics that capture discriminatory features separating $\Xrel$ from $\Xirr$. Specifically, for each $\bx_j$, we compute the sample mean of $\bm{c}_j$ and $\bm{R}_j^c$, 25th percentile of $\bm{c}_j$, and 75th percentile of $\bm{R}_j^c$, yielding the summary vector $\bm{Z}_j = \left(\bar{c}_j, \mathcal{Q}_{0.25}(\bm{c}_j), \overline{R}_j^c, \mathcal{Q}_{0.75}(\bm{R}_j^c)\right)$. By stacking these summary vectors row-wise, the resulting VC-measure summary matrix 
\begin{equation}\label{eq:summary_matrix}
    \bm{Z} = (\bm{Z}_1^T, \ldots, \bm{Z}_p^T)^T \in \mathbb{R}^{p \times 4}
\end{equation}
serves as the clustering feature matrix whose columns offer complementary discriminatory signals. The use of sample mean is motivated by the rank characterization \eqref{eq:vip_rank_assumption} in Section~\ref{sec:cluster_vip_rank}, under which relevant predictors tend to have much higher average ranks than irrelevant ones. We include $\mathcal{Q}_{0.25}(\bm{c}_j)$ to capture the lower end of a variable's splitting frequency, since relevant predictors tend to have high VC values throughout all runs, not just occasionally. Conversely, $\mathcal{Q}_{0.75}(\bm{R}_j^c)$ helps ensure that $\Xirr$, which may spike toward high-ranking positions (i.e., $\{1,\ldots,p_0\}$) occasionally, are still distinguished by their poor rankings (i.e., $\{p_0+1,\ldots,p\}$) in most runs. Together, these statistics provide a more robust separation between $\Xrel$ and $\Xirr$ under both high- and low-noise regimes. To reduce skewness and stabilize within-cluster variance, we apply a \texttt{log1p} transformation (i.e., $\log(1+x)$) to each column of $\bm{Z}$ \citep{BoxCox1964}. Each column is then standardized to have zero mean and unit variance, ensuring equal contribution to clustering \citep{ESL}. If prior knowledge is available about the informativeness of each statistic, one may optionally assign feature-specific weights. The pseudo-code for the VC-measure is summarized in Section A of Supplementary Material. 

Since VC and VC Rank can be easily computed from posterior samples of \emph{any} BART variant, the VC-measure matrix $\bm{Z}$ in \eqref{eq:summary_matrix} is broadly applicable. Throughout the paper, we use the naming convention ``XXX VC-measure'' to denote a variable selection method whose summary matrix $\bm{Z}$ is derived from the posterior samples of the BART variant ``XXX.'' Similarly, the summary matrix $\bm{Z}$ can also be constructed using VIP and VIP Rank, and we refer to such methods as ``XXX VIP-measure.'' In particular, \vipRank is a special case of BART VIP-measure, where the summary matrix $\bm{Z}$ only consists of a single column of average VIP Ranks $\left(\overline{R}_1^q,\ldots,\overline{R}_p^q\right)$.

\section{Numerical Experiments}
\label{sec:sim}

In this section, we present a comprehensive simulation study comparing the proposed VC-measure with seven existing Bayesian tree ensemble variable selection methods using 3,600 scenarios, each replicated 10 times. 

Specifically, the seven existing methods are \Local, \Gse, and \Gmax, \MI, and \vipRank under the general BART prior; \dart and \abc  under regularization Bayesian tree priors. The proposed posterior summary is paired with BART and DART, leading to two new variable selection methods, ``BART VC-measure'' and ``DART VC-measure.''

For the four permutation-based methods (\Local, \Gse, \Gmax, and \MI), we set $\Lperm = 50$ and $\alpha = 0.05$. In DART, the Dirichlet hyperprior parameter $\alpha$ in \eqref{eq:s_prior} is drawn from $\alpha/(\alpha + p) \sim \text{Beta}(0.5, 1)$. \abc use default settings: the spike-and-slab prior $\pi(\mathcal{S})$ follows a beta-binomial distribution with $\theta_0 \sim \text{Beta}(1,1)$; data are randomly split 50/50 between prior training and ABC acceptance/rejection; and the top 10\% of $\Labc = 1,000$ ABC samples are used to estimate MPVIP. All methods use $\Lrep = 10$ independent fits to stabilize the variable importance measure, unless stated otherwise; exceptions include the original DART where one fit is used, \abc where $\Labc = 1,000$ is used, and \vipRank where $\Lrep = 20$ is used. All methods are trained with $T = 20$ trees to promote predictor competition. A burn-in of 5,000 posterior samples followed by 5,000 additional draws is used for all methods except \abc, which instead follow \cite{ABC_forest}: 200 burn-in samples, and the 201st posterior sample is used for MPVIP estimation. All other hyperparameters remain at their default values. The methods are implemented using their respective R packages: \texttt{bartMachine v1.3.4.1} (\Local, \Gse, and \Gmax), \texttt{BartMixVs v1.0.0} (\MI), \texttt{dartMachine v1.2.0} (DART), and \texttt{BartVC v1.0.0} (clustering-based methods). For the two new methods, BART VC-measure and DART VC-measure, we follow common practice in the BART literature and independently train the models $\Lrep = 10$ times as described above, although such repeated fits are not strictly necessary: reducing $\Lrep$ to 2 in DART VC-measure still outperforms all competing methods (Section~\ref{sec:diff_L}).

\begin{table*}[t]
\caption{Settings used in the experiments.} \label{tab:setting}
\vskip 0.1in
\begin{center}
\begin{sc}
\begin{tabular}{lll}
\toprule
Setting  & Value\\
\midrule
\# of datasets           & 100 \\
\# of trials per dataset & 10  \\
\# of methods            & 8  \\
$n$                      & 500, 1000, 1500, 2000 \\
SNR                      & 0.25, 0.5, 1, 2, 5, 10, 15, 20, noiseless \\
$p_0$                    & 2 to 9 \\
$p$                      & 102 to 459  \\
\# of data settings      & 3,600 \\
\# of comparisons        & 288,000 \\
\bottomrule
\end{tabular}
\end{sc}
\end{center}
\end{table*}

The experiment settings are summarized in Table~\ref{tab:setting}. We evaluate the aforementioned methods using the high-dimensional Feynman database introduced in \cite{pan_sr}, totaling 100 nonlinear physics equations from the \textit{Feynman Lectures on Physics} \citep{Feynman_Lectures}. Specifically, for each equation $f_0(\cdot)$ in the \textit{Feynman Lectures on Physics}, we generate the relevant features $\Xrel$ as $(x_{1,j},\ldots,x_{n,j}) \overset{\text{iid}}{\sim} \text{Unif}(a_j, b_j)$ for $1 \leq j \leq p_0$, where $p_0 = |\mathcal{S}_0|$, $n$ is the sample size, and $(a_j,b_j)$ are the lower and upper bounds for feature $\bx_j$ described in \cite{AIFeynman1.0}. Then, the response variable is generated as $y_i = f_0(x_{i,1},\ldots,x_{i,p_0}) + \varepsilon_i$ with $\varepsilon_i \overset{\text{iid}}{\sim} \mathcal{N}(0, \sigma_\varepsilon^2)$ for $1 \leq i \leq n$, where $\sigma_\varepsilon^2 = \sigma_f^2/\text{SNR}$ is specified through the SNR and the true variance $\sigma_f^2 = \text{Var}(f_0)$.

In addition to the relevant features $\bX_{\mS_0} = (\bx_1,\ldots,\bx_{p_0})$, we include an array of irrelevant features $\bX_{\text{irr}}$, representing the era of big data where all reasonable features are included in the dataset. Specifically, for each relevant feature $\bx_j$,  $j \in \mS_0$, we generate $(\bx_{j,\text{irr}}^1, \ldots, \bx_{j,\text{irr}}^S) \overset{\text{iid}}{\sim} f_{x_j}$, representing $S$ copies of independent and irrelevant features coming from the same distribution as $\bx_j$. Then, the final feature matrix is $\bX = [\bX_{\mS_0}, \bX_{\text{irr}}^1,\ldots,\bX_{\text{irr}}^{p_0}] \in \mathbb{R}^{n \times p}$, where $\bX_{\text{irr}}^j = (\bx_{j,\text{irr}}^1, \ldots, \bx_{j,\text{irr}}^S) \in \mathbb{R}^{n \times S}$ is the irrelevant feature matrix induced by the $j$th relevant feature for $j = 1,\ldots,p_0$, and $p = p_0(1+S)$ is the total number of features. In the following, we fix $S=50$ so the total number of features is $p=51p_0$.

To evaluate their robustness to noise and sample size, all methods are evaluated under the 3,600 scenarios formed by the 100 Feynman equations, $n \in \{500, 1000, 1500, 2000\}$, and $\text{SNR} \in \{0.25, 0.5, 1, 2, 5, 10, 15, 20, \text{noiseless}\}$, and each scenario is replicated 10 times, totaling 36,000 fits for each method.

\subsection{Performance Metrics}\label{sec:metrics}
We assess the variable selection accuracy using true positive rate (TPR), false positive rate (FPR), and $F_1$ score, given by
$$\text{TPR} = \frac{\text{TP}}{\text{TP}+\text{FN}}, \quad \text{FPR} = \frac{\text{FP}}{\text{FP} + \text{TN}}, \quad F_1 = \frac{2\text{TP}}{2\text{TP} + \text{FP} + \text{FN}},$$
where TP is the number of relevant predictors correctly identified, FP is the number of irrelevant predictors selected, FN is the number of relevant predictors not identified, and TN is the number of irrelevant predictors not selected. A higher TPR (or recall) is desirable as it indicates that most of the relevant predictors $\Xrel$ are selected. On the contrary, a lower FPR is favorable as it indicates that irrelevant predictors $\Xirr$ are rarely selected. The $F_1$ score is a harmonic mean of precision and recall, which balances a procedure's capability to make necessary selections with its ability to avoid including irrelevant predictors. When $F_1 = 1$, the set of selected variables $\widehat{\mathcal{S}}$ coincides with the ground-truth $\mathcal{S}_0$. Although infrequent, ``no selection'' and ``does not complete'' do happen. In such scenarios, we set $\text{TPR} = \text{FPR} = F_1 = 0$. 

In addition to selection accuracy, we also evaluate the computational cost of each method. Specifically, all simulations are trained on the same server with an AMD EPYC 7642 CPU, where each job is limited to 1 CPU core and no parallelization is allowed. Runtime in seconds across different settings of $(n,p)$ is compared in Section~\ref{sec:runtime}. 



\begin{table}[t]
    \caption{Variable importance measures for BART and its variants.} \label{tab:measures}
    \vskip 0.1in
    \centering
    \begin{tabular}{@{}lcc@{}}
    \toprule
    \textbf{Importance Measure} & \textbf{Selection Criterion} & \textbf{Prior} \\
    \midrule
    VIP                         & Permutation with $\Lperm=50$ & BART \\
    \midrule
    VIP Rank                    & Clustering                   & BART \\
    \midrule
    MI                          & Permutation with $\Lperm=50$ & BART \\
    \midrule
    MPVIP                       & Median Probability Model (MPM)                          & \makecell{ABC Bayesian forests \\ DART} \\
    \midrule
    VC-measure                  & Clustering                   & All \\
    \bottomrule
    \end{tabular}
\end{table}

\subsection{Superiority of VC-measure Compared with Existing Posterior Summaries for Each Bayesian Tree Prior}
\label{sec:superior_of_VC}
The VC-measure can be used as a drop-in replacement across Bayesian tree models. Table~\ref{tab:measures} summarizes both existing and proposed variable importance measures for BART and its variants: VIP, VIP Rank, and MI are tailored to the original BART prior, while MPVIP is designed for ABC Bayesian forests and DART. All results below are averaged over 100 equations, each replicated 10 times. 

We begin by pairing VC-measure with the BART prior in place of VIP, VIP Rank, and MI, defining a new method, \emph{BART VC-measure}, with $\Lrep = 10$ matching VIP and MI and all other hyperparameters as in Section~\ref{sec:sim}. To streamline comparison, we report results only for the best performing VIP-based method, BART VIP-G.SE, based on $F_1$ score; full comparisons among VIP-G.SE, VIP-G.Max, and VIP-Local are provided in Sections~\ref{sec:comparison} and \ref{sec:FP.TP}.

\begin{figure}[t]
    \centering
    \includegraphics[width=\linewidth]{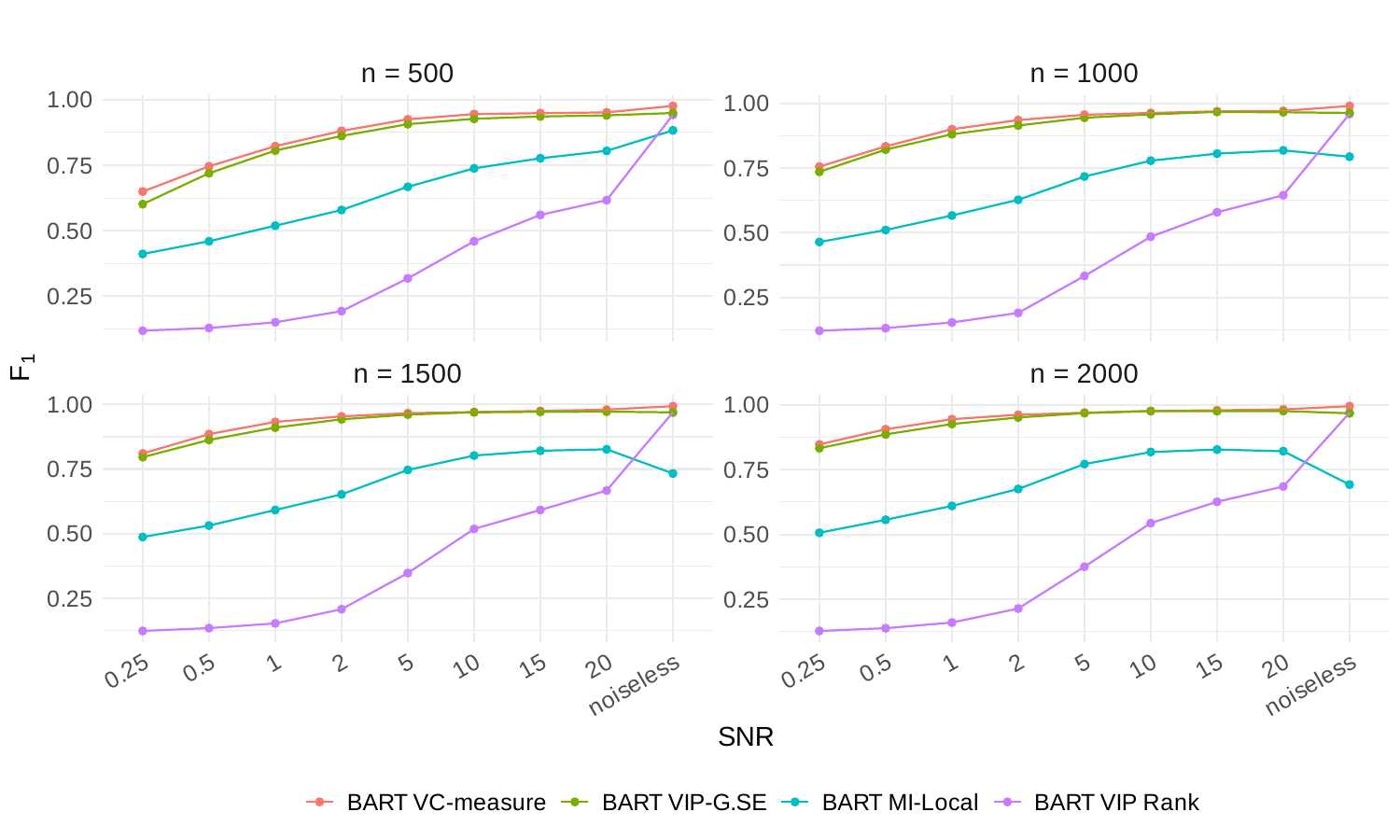}
    \caption{$F_1$ score comparison among variable importance measures for BART prior. Points indicate the average $F_1$ scores over 100 Feynman equations, each with 10 replicates.}
    \label{fig:BART_L10}
\end{figure}

Figure~\ref{fig:BART_L10} shows that substituting VC-measure for existing measures uniformly improves $F_1$ score across all simulation settings. BART VIP-G.SE performs competitively in large-$n$ scenarios, but at a substantially higher computational cost, approximately six times that of BART VC-measure due to the additional $\Lperm = 50$ permutation runs. BART MI-Local ranks third in most cases but underperforms in noiseless settings, where it occasionally fails to select any variables, resulting in an $F_1$ score of 0. BART VIP Rank has the lowest $F_1$ score by a large margin, reflecting its emphasis on high true positive rates (TPR) at the expense of increased false positives (FPs); this is a recurring trade-off discussed in Sections~\ref{sec:cluster_vip_rank}, \ref{sec:comparison}, and \ref{sec:FP.TP}. 

\begin{figure}[t]
    \centering
    \includegraphics[width=\linewidth]{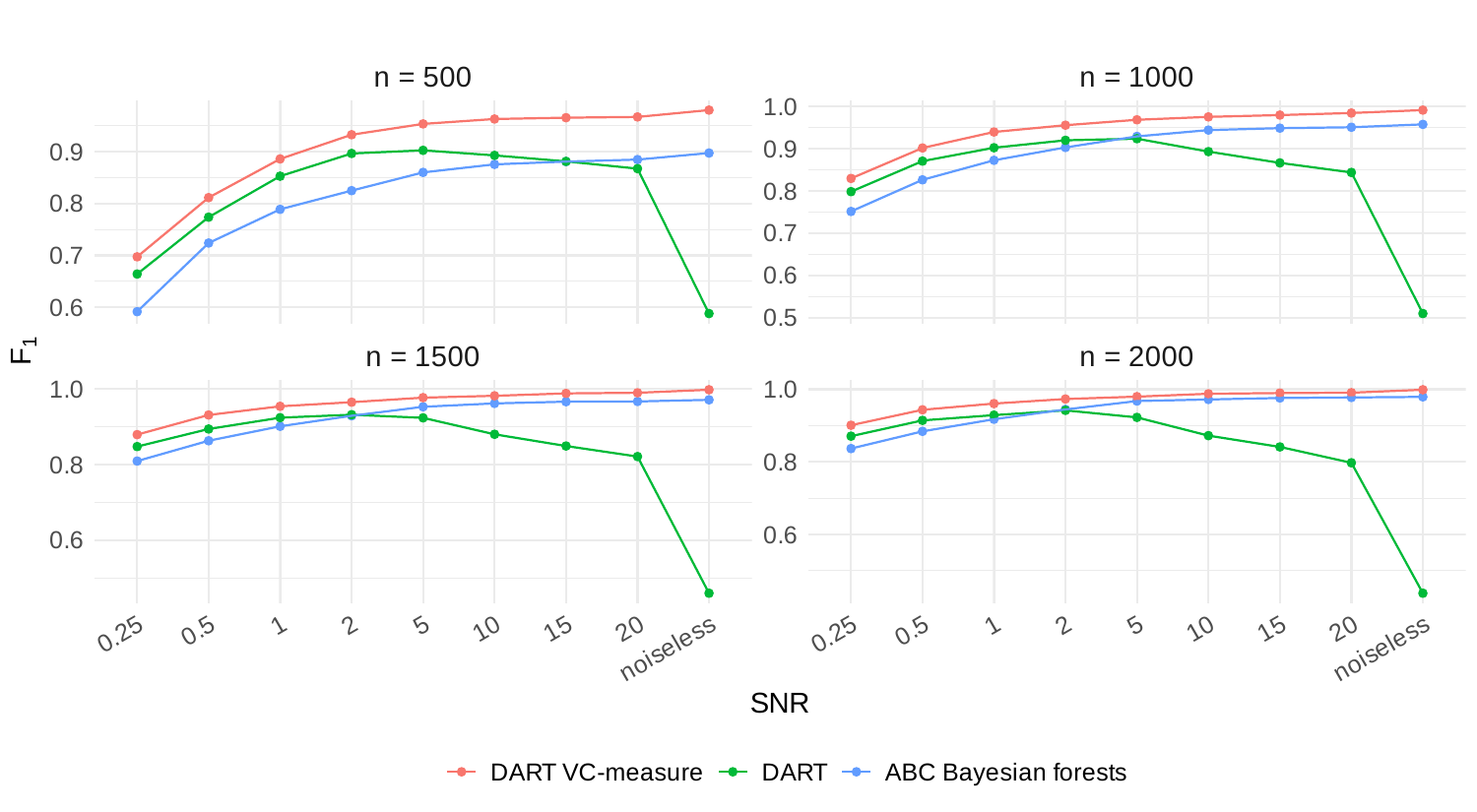}
    \caption{$F_1$ score comparison among MPVIP-suitable priors. Points indicate the average $F_1$ scores over 100 Feynman equations, each with 10 replicates.}
    \label{fig:DART_L10}
\end{figure}

We next evaluate VC-measure as a replacement for MPVIP. Among the two MPVIP-suitable priors, we pair VC-measure with the DART prior and refer to this new method as \emph{DART VC-measure}. As shown in Figure~\ref{fig:DART_L10}, VC-measure vastly improves over the original DART across all simulation settings. Moreover, it resolves the selection instability observed in DART when $\text{SNR} \geq 5$. We show in Section~\ref{sec:FP.TP} that this instability is attributed to increased FPs and defer its root cause to that section. In addition to outperforming DART, DART VC-measure consistently surpasses ABC Bayesian forests in $F_1$ score, especially in small-$n$ regimes, where ABC Bayesian forests are disadvantaged by their 50/50 training/proposal data split, resulting in fewer samples for training. A direct comparison of BART and DART on the same VC-measure posterior summary is given in Section B of Supplementary Materials.

In summary, VC-measure serves as a robust drop-in replacement for existing variable importance measures across the BART family, achieving uniformly higher $F_1$ scores than the original methods with significant computational advantages over permutation-based approaches, and its performance remains stable across a wide range of $(n, \text{SNR})$ configurations.

\subsection{Superiority of DART VC-measure Compared with All Other Methods}\label{sec:comparison}

\begin{figure}[t]
    \centering
    \includegraphics[width=0.97\linewidth]{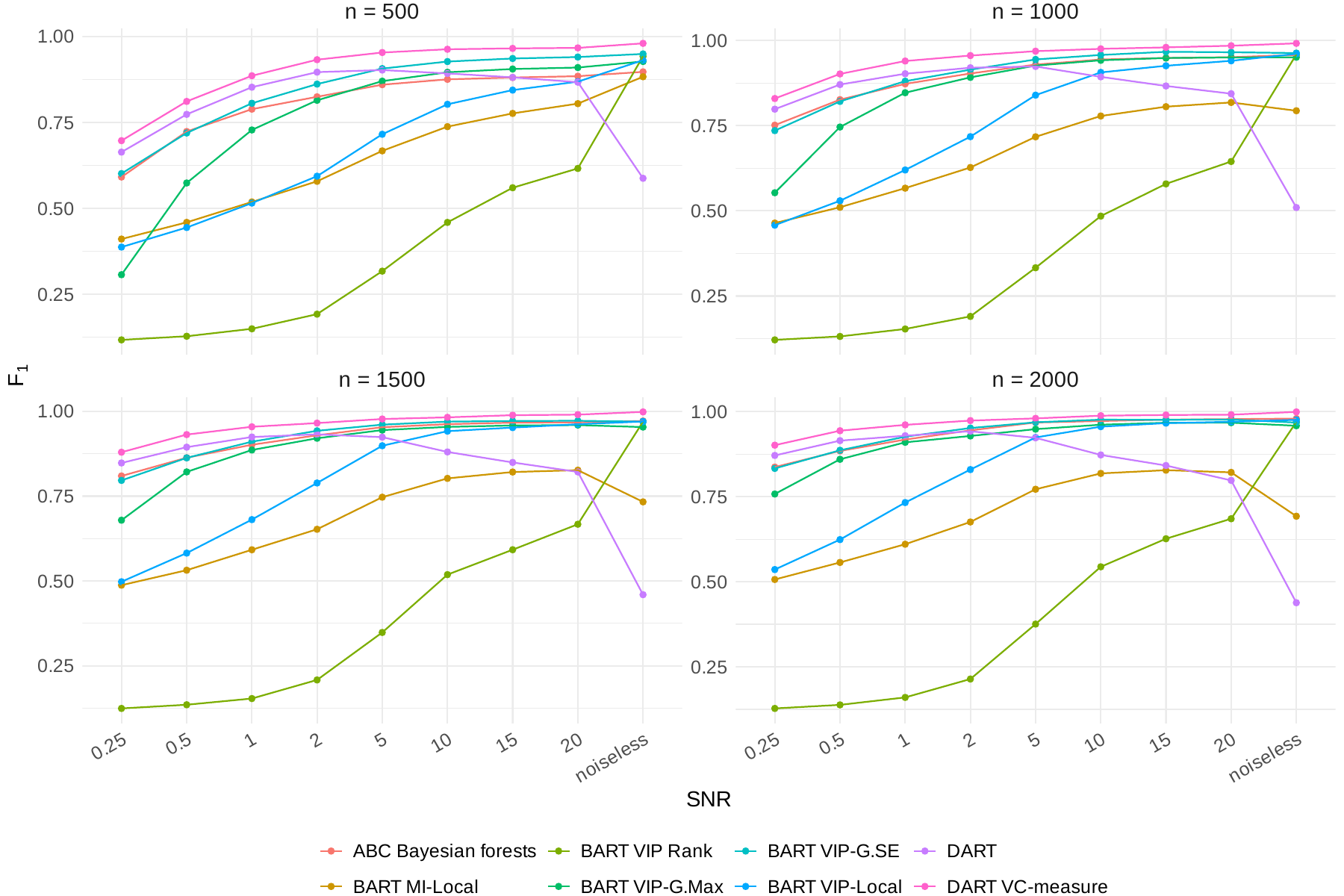}
    \caption{$F_1$ score. Points indicate the average $F_1$ score over 100 Feynman equations, each with 10 replicates.}
    \label{fig:f1}
\end{figure}
In this section, we compare all methods described in Section~\ref{sec:sim} using the high-dimensional Feynman database. As shown in Figure~\ref{fig:f1}, the proposed method, \vcMeasure with $\Lrep = 10$, consistently outperforms all other methods in terms of $F_1$ score, demonstrating robustness across all combinations of $(n, \text{SNR})$. Notably, in noisy settings $(\text{SNR} \leq 1)$, the original DART algorithm ranks second. However, as SNR increases, its $F_1$ score deteriorates due to frequent false positive (FP) selections (see Figure~\ref{fig:fpr}). Although the Dirichlet hyperprior in \dart down-weights the splitting probabilities for $\Xirr$, it does not eliminate them entirely. In particular, a subset of $\Xirr$ is found to have low split frequencies (typically 1 to 3 splits within an ensemble) persistently over 50\% of the posterior draws, thereby inflating their MPVIPs over 0.5 and leading to false discoveries. In contrast, \vcMeasure avoids this pitfall despite using the same posterior samples. Unlike MPVIP, where having $\text{VC}=1$ in $\geq 50\%$ of posterior samples is sufficient to be deemed ``important,'' VC-measure distinguishes signal from noise based on the magnitude of VC \emph{and} the persistence of large VC values over model replications. As such, a variable with a consistent low frequency split (such as having $\text{VC}=1$ in many posterior samples) is highly unlikely to be selected in VC-measure. \abc, which also use MPVIP with the MPM criterion, do not suffer the same issue: their spike-and-slab prior directly controls the set $\mathcal{S}$ of predictors available for splitting, and owing to the variable selection consistency theory established in \cite{ABC_forest}, $\mathcal{S}$ rarely contains irrelevant predictors $\Xirr$, so splits on them are rare and impersistent over different ABC iterations.

Among the permutation-based methods, \Gse performs the best in all but the noiseless settings with $n \geq 1500$, where \Local is marginally higher, achieving a strong balance between precision and recall. \Gmax ranks closely behind, especially in low-noise settings $(\text{SNR} \geq 2)$, where its $F_1$ scores nearly match those of \Gse. However, its performance degrades in noisier settings; for example, at $n=500$ and $\text{SNR} = 0.25$, it ranks second to last among all methods due to its stringent selection threshold. Unsurprisingly, \Local and \MI perform worse, reflecting their lenient inclusion thresholds that favor FPs over FNs. As a result, these methods can safely identify most, if not all, relevant predictors $\Xrel$ (Figure~\ref{fig:tpr}), making them suitable for tasks such as pre-screening and exploratory analysis. Among the two, \Local outperforms \MI except in the noisiest settings ($\text{SNR} \leq 1$ for $n = 500$ and $\text{SNR} = 0.25$ for $n = 1000$), and the $F_1$ scores of \MI drop sharply in the noiseless, larger sample size $(n = 1000, 1500, 2000)$ settings due to occasional ``no selection,'' in which case we assign an $F_1$ score of 0. By taking the philosophy of favoring FP over FN to the extreme, \vipRank often ranks last, with a substantially lower $F_1$ than the next-best method. 

\begin{figure}[t]
    \centering
    \includegraphics[width=\linewidth]{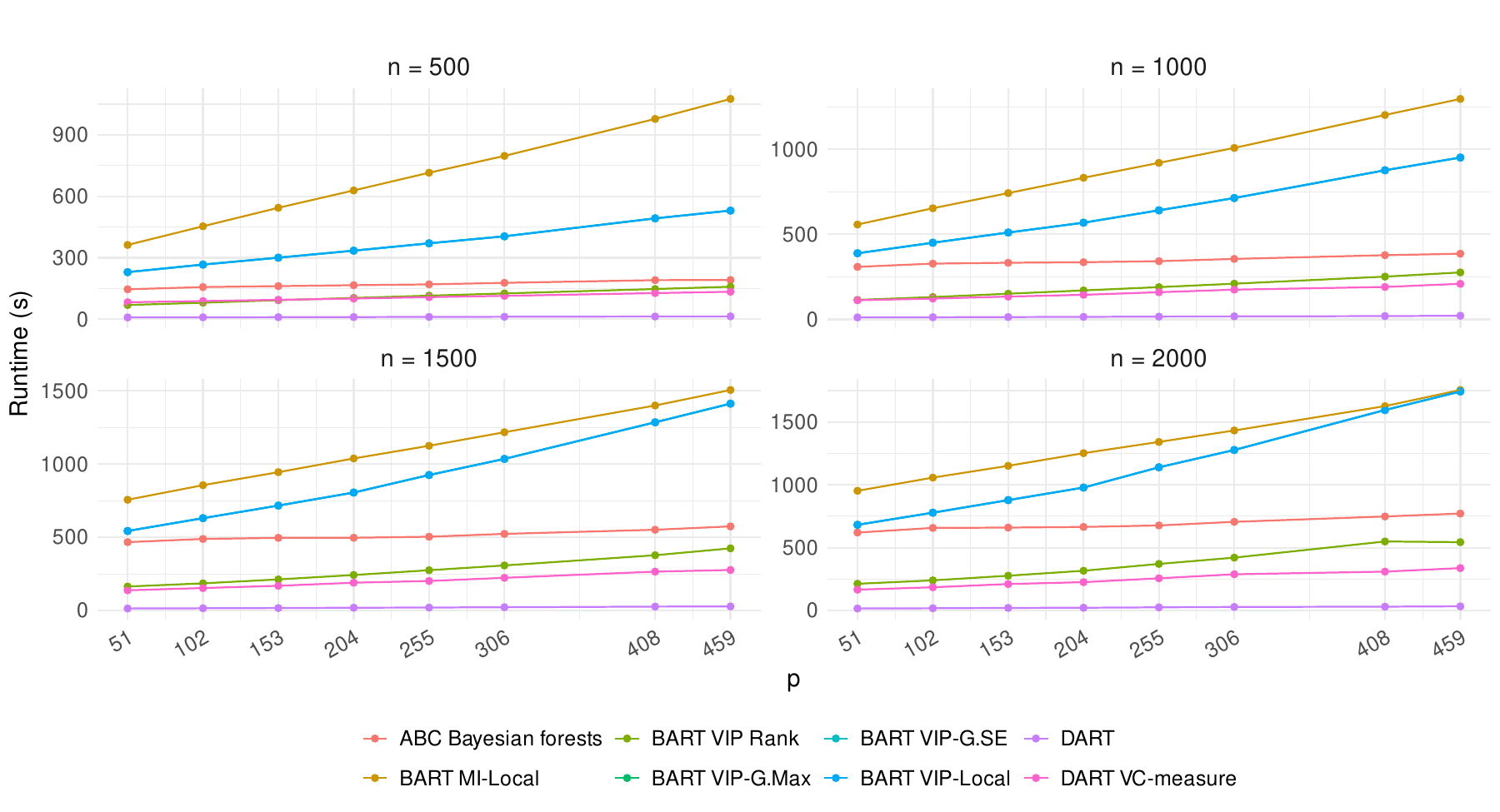}
    \caption{Runtime in seconds. Points indicate the average runtime over 100 Feynman equations, each with 10 replicates. DART VC-measure can be further sped up 5-fold by using $\Lrep = 2$ without sacrificing accuracy, as described in Section~\ref{sec:diff_L}.}
    \label{fig:runtime}
\end{figure}

\subsection{Computational Cost}
\label{sec:runtime}

Figure~\ref{fig:runtime} shows the average runtime (in seconds) of the seven existing methods and DART VC-measure as a function of the number of input features $p$, since runtime is largely independent of SNR. The fastest method by far is \dart, which requires only a single fit per simulation (10,000 posterior draws in total) and completes in under 34 seconds on average even in the most challenging setting $(n,p) = (2000, 459)$. The clustering-based methods are also efficient, with \vcMeasure coming in second (100,000 posterior draws) and \vipRank in third (200,000 posterior draws). \vcMeasure's computational cost is essentially 10 times that of \dart due to the 9 additional model replications, and, as discussed in Section~\ref{sec:diff_L}, reducing $\Lrep$ to 2 drastically reduces its runtime while maintaining a superior $F_1$ score in all simulation settings. Although \vipRank requires $\Lrep = 20$, double that of \vcMeasure, its runtime is almost always less than twice that of \vcMeasure since DART requires additional Gibbs updates for the variable splitting probabilities $\bm{s}$. The computational cost of \abc is mainly attributed to the number of ABC iterations $\Labc = 1,000$, each of which corresponds to a small BART model fit with a burn-in period of 200 draws followed by 1 posterior draw; hence, \abc require $201 \Labc = 201,000$ posterior draws per simulation, each with more parameter updates than under the BART and DART priors, which amounts to fourth place overall. Permutation-based methods rank last in terms of efficiency since each requires $\Lrep=10$ model fits on the original data for a robust VIP estimation and additional $\Lperm = 50$ fits on permuted data to estimate the null VIP distribution, amounting to 600,000 posterior draws per simulation, 6 times that of \vcMeasure and 60 times that of \dart. Furthermore, \MI is consistently slower than \Local, likely due to a combination of increased arithmetic in Metropolis importance (MI) calculations and differences in implementation (\MI in C++ vs. \Local in Java).

\subsection{Sensitivity of \vcMeasure to $\Lrep$}\label{sec:diff_L}

\begin{figure}[t]
    \centering
    \includegraphics[width=0.9\linewidth]{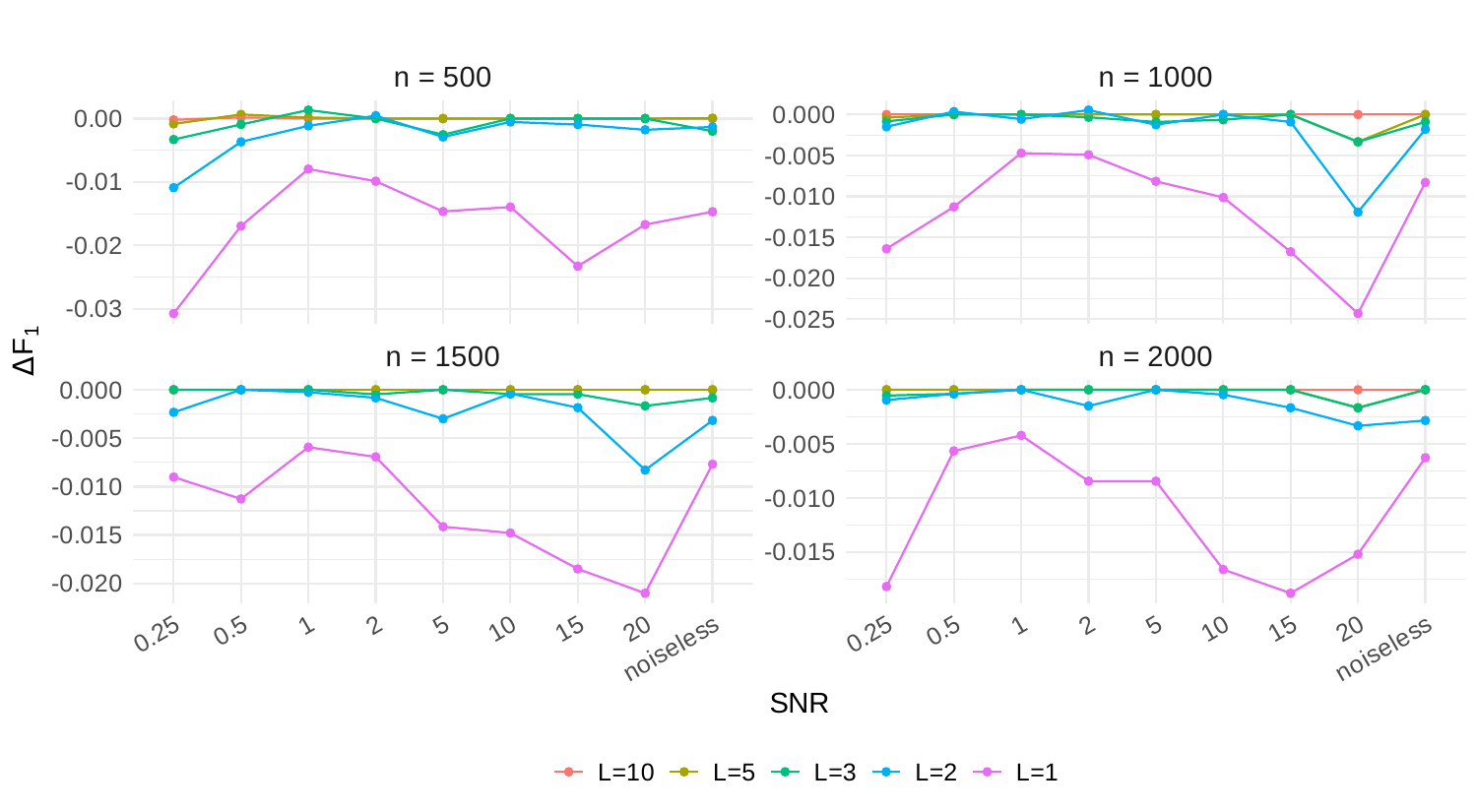}
    \caption{Differences in $F_1$ score compared to \vcMeasure with $\Lrep = 20$ model replications. Points indicate the average differences in $F_1$ score over 100 Feynman equations, each with 10 replicates.}
    \label{fig:vc_diff_L}
\end{figure}

We assess the sensitivity of \vcMeasure to its sole tuning parameter, the number of model replications $\Lrep$. For each of the 3,600 simulation settings in Table~\ref{tab:setting}, each repeated 10 times, we train 20 independent DART models and compute VC-measure using the first $\Lrep$ models, where $\Lrep \in \{1,2,3,5,10,20\}$. For reference, we define the \emph{Next Best} pseudo competitor as the best $F_1$ score among all non-VC-measure methods in the same simulation setting, which allows us to evaluate how small $\Lrep$ can be while still surpassing all alternatives.

When $\Lrep = 1$, the VC Rank matrix $\bm{R}^c$ reduces to a single permutation of the variable indices $\{1,\ldots,p\}$, providing no discriminatory power, and the VC mean and 25th percentile reduce to the raw VC value $c_j$; thus, the summary matrix $\bm{Z}$ carries no information beyond the single-fit VC vector $(c_1, \ldots, c_p) \in \mathbb{R}^p$.

Let $F_1(\Lrep)$ denote the average $F_1$ score of \vcMeasure with $\Lrep$ replications, and define $\Delta F_1(\Lrep) = F_1(\Lrep) - F_1(20)$, where $\Lrep = 20$ serves as the baseline. Then, $\Delta F_1(\Lrep) = 0$ indicates identical $F_1$ score to the baseline model. Figure~\ref{fig:vc_diff_L} shows that for $\Lrep \in \{5, 10\}$, $\Delta F_1(\Lrep)$ is zero in all but a handful of settings and never below $-0.0033$, indicating no practical loss in accuracy while reducing computational cost by at least 50\%. 

Lower values $\Lrep \in \{1,2,3\}$ yield slightly more variable results, as fewer replications provide less information about the variability of VC across model fits. This variability is crucial for the summary statistics vector $\bm{Z}_j = \left(\bar{c}_j, \mathcal{Q}_{0.25}(\bm{c}_j), \overline{R}_j^c, \mathcal{Q}_{0.75}(\bm{R}_j^c)\right)$ to provide discriminatory power to distinguish relevant predictors from the irrelevant ones. Nonetheless, performance remains competitive. In particular, $\Lrep = 2$ achieves $\Delta F_1(2) \geq -0.012$ across all settings and still exceeds the \emph{Next Best} competitor in every setting, by at least $0.006$ in $F_1$. Even with $\Lrep = 1$, \vcMeasure remains comparable to state-of-the-art methods, with $\Delta F_1(1) \geq -0.031$ in all settings.

Overall, Figure~\ref{fig:vc_diff_L} demonstrates that \vcMeasure is not only superior to existing methods (Section~\ref{sec:comparison}), but also remarkably robust to the choice of $\Lrep$. We recommend $\Lrep = 10$ as a default, balancing accuracy and stability while achieving $\Lrep = 20$ level of performance in all settings. With parallel computing, the wall-clock time for $\Lrep$ replications can approach that of a single model fit, making \vcMeasure as fast as DART, the fastest method tested. When the computational budget is limited, $\Lrep = 2$ offers an attractive trade-off, requiring only twice the runtime of DART while consistently outperforming all other methods, many of which are 10-76 times slower than DART.

\subsection{Comparison of False Positives and True Positives}
\label{sec:FP.TP}

\begin{figure}[t]
    \centering
    \includegraphics[width=\linewidth]{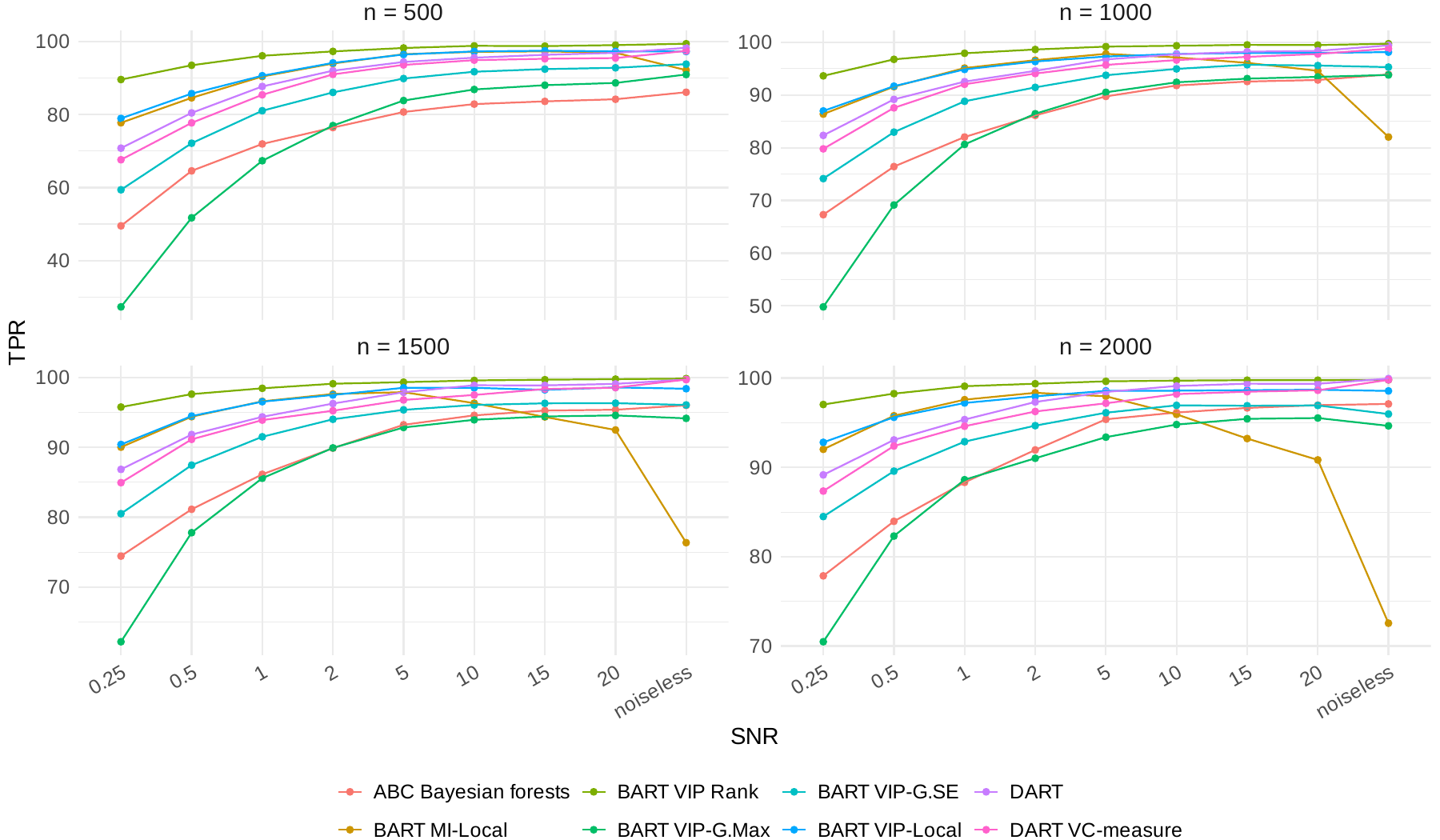}
    \caption{True positive rate (TPR). Points indicate the average TPR over 100 Feynman equations, each with 10 replicates.}
    \label{fig:tpr}
\end{figure}

While the $F_1$ score balances false positives and true positives, certain applications weigh them asymmetrically. To provide additional insight into the characteristics of different methods, we evaluate them separately on TPR and FPR in this section.

\begin{figure}[t]
    \centering
    \includegraphics[width=\linewidth]{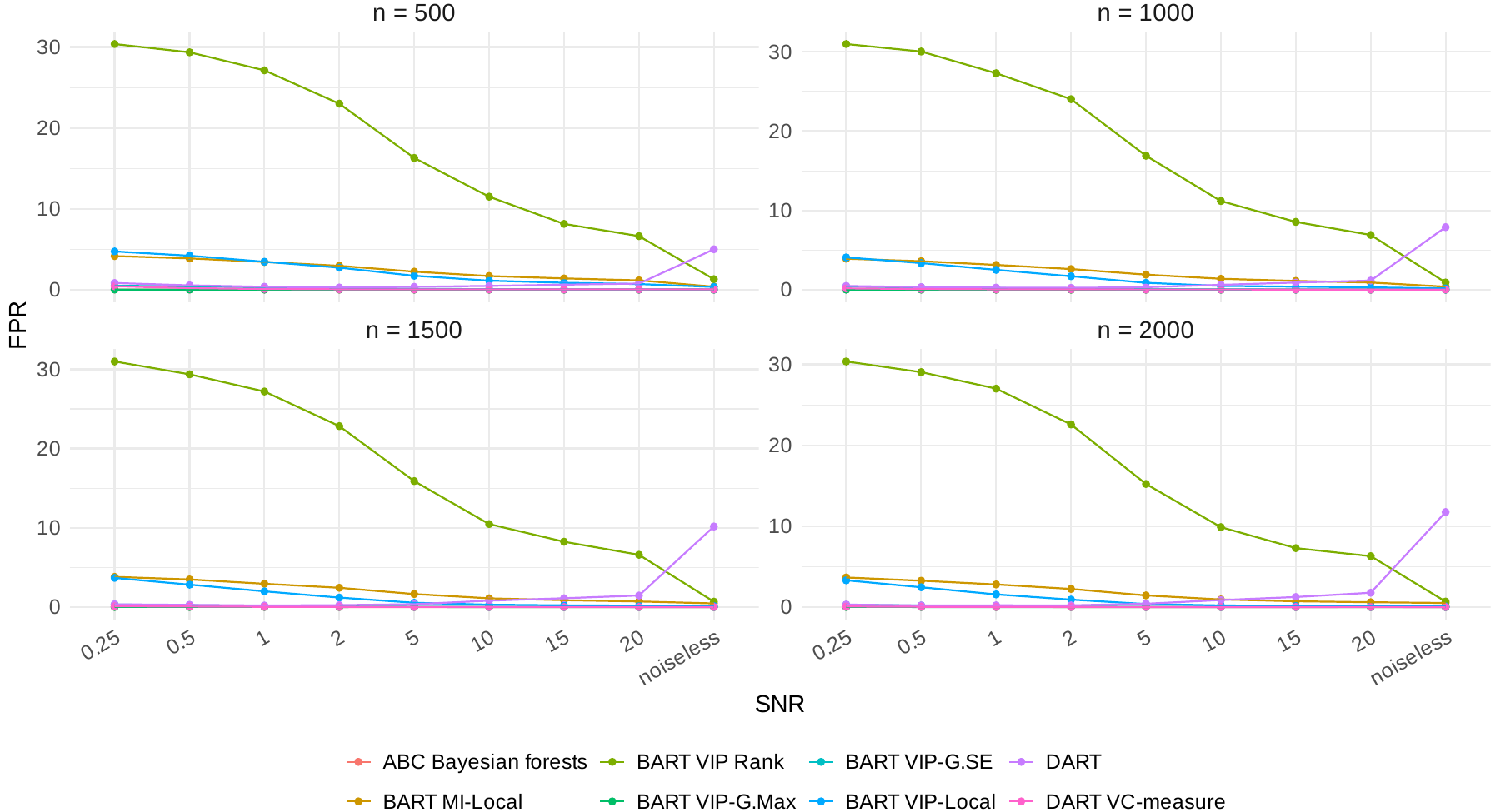}
    \caption{False positive rate (FPR). Points indicate the average FPR over 100 Feynman equations, each with 10 replicates.}
    \label{fig:fpr}
\end{figure}

In terms of TPR (Figure~\ref{fig:tpr}), \vipRank consistently achieves the highest TPR, with $\text{TPR}\geq90\%$ across all simulation settings. Similarly, the less stringent methods, \Local and \MI, also show strong TPR performance, though \MI occasionally selects no variables in high-SNR, large-$n$ settings, resulting in notable TPR drops. The two DART-based methods, \dart and \vcMeasure, rank fourth and fifth in most settings, with \dart slightly outperforming \vcMeasure. Interestingly, in low noise $(\text{SNR} \geq 10)$ and large sample size ($n \geq 1000)$ settings, both DART-based methods achieve TPRs comparable to \Local, and even surpass it when $n=2000$. \Gse closely follows the DART-based methods, typically showing a 2-5\% lower TPR. Owing to their stringent selection criteria, \abc and \Gmax consistently rank lowest in TPR. While \abc marginally outperform \Gmax in most scenarios, this trend reverses at $n = 500$ and $\text{SNR} \geq 2$. Notably, \Gmax performs especially poorly when $\text{SNR}=0.25$, with an average TPR of only 27.3\% at $(n, \text{SNR}) = (500, 0.25)$, highlighting its unsuitability for pre-screening tasks where capturing all relevant predictors $\Xrel$ is crucial for downstream analysis and modeling.

Figure~\ref{fig:fpr} illustrates the FPR comparison across all eight methods. Only methods with a strong preference for FP over FN, namely \vipRank, \Local, and \MI, exhibit non-negligible FPRs. Among them, \vipRank stands out with particularly high FPRs, even exceeding the next highest methods, \Local and \MI, by a considerable margin. At $\text{SNR} \leq 10$, \vipRank's FPR ranges from 10-30\%, reflecting its lenient variable inclusion strategy. While this may be acceptable for exploratory and pre-screening tasks, it makes \vipRank poorly suited for settings where validation is costly, such as gene knockout experiments. \Local and \MI also show elevated FPR (typically below 5\%), reflecting their design to prioritize inclusion of all relevant features, even at the cost of FPs. In contrast, the remaining five methods maintain a near-zero FPR across all settings, though this comes at the expense of low TPR. Interestingly, \dart shows a noticeable increase in FPR at higher SNR levels, corresponding with its drop in $F_1$ score discussed earlier. This pattern supports the idea that in low-noise settings, \dart's MPVIPs become prone to \emph{overconfident inclusion} of irrelevant predictors. 

These differences highlight a fundamental trade-off between inclusion (high TPR) and exclusion (low FPR), so the choice of method should be guided by the downstream application: in exploratory settings, such as predictor pre-screening for scaling symbolic regression \citep{pan_sr,iBART}, \vipRank or \Local may be preferred despite their higher FPRs, whereas in confirmatory or resource-intensive settings, such as experimental design or clinical validation, more conservative methods like \Gse or \abc are more appropriate.

\section{Discussion}
\label{sec:dis}

Variable selection with Bayesian tree ensembles is usually approached by designing the tree prior, whereas the posterior summary that turns posterior samples into a selected set has received far less attention. This paper shows that the choice of posterior summary can matter as much as the choice of prior. The proposed VC-measure, which summarizes raw variable counts and their ranks over a few independent fits using four statistics and applies clustering-based thresholding, is simple, tuning-free, and applicable to any BART variant. With the same posterior samples, it uniformly improves the $F_1$ score of both general BART and DART across a benchmark of 3,600 settings spanning a wide range of SNR, sample sizes $n$, and data-generating functions, and, in all but the four noisiest settings with $n \leq 1000$, its gain over the analogous VIP-based summary exceeds that of switching from BART to DART under the same summary (Section C of Supplementary Material). Paired with DART, it consistently achieves the highest $F_1$ scores across varying $n$ and SNR, and the gain persists with as few as $\Lrep = 2$ fits, so that \vcMeasure adds little computational burden over a standard DART run.

The benchmark also clarifies why existing summaries fall short and when they remain appropriate. The original DART algorithm, which pairs MPVIP with the MPM threshold, is competitive in noisy settings but loses accuracy as SNR increases due to increased false positives. We hypothesize that this is caused by posterior overconcentration, where occasional splits on irrelevant variables $\Xirr$ inflate their MPVIPs, causing spurious selections; this behavior persists even when $\sigma^2$ is fixed at its oracle value, suggesting that the Dirichlet prior alone is insufficient to suppress these FPs in low-noise regimes. To our knowledge, this failure mode of MPVIP-based selection in concentrated posteriors has not been previously documented and warrants further theoretical investigation. The VC-measure avoids this pitfall with the same posterior samples because it weighs the magnitude and persistence of variable counts rather than their mere occurrence. Permutation-based methods, especially \Gse, are accurate and stable, ranking second or third in $F_1$ in most settings, but require $\Lrep + \Lperm$ model fits, which limits their use in high-throughput or time-constrained applications; \abc are computationally lighter than permutation-based methods but frequently exclude relevant variables due to their stringent spike-and-slab prior, consistent with the findings in \cite{BartMixVs}. Methods with lenient inclusion thresholds, such as \Local, \MI, and \vipRank, achieve the highest TPR, often exceeding 90\% when $n \geq 1000$, at the cost of a higher FPR, which can exceed 30\% for \vipRank in low-SNR settings.

These trade-offs underscore the importance of aligning the choice of method with the downstream application. When computational budget is tight but variable selection accuracy remains critical, \vcMeasure with a small $\Lrep$ (e.g., 1 or 2) offers a strong balance of performance and efficiency. When the goal is exploratory screening with high recall, \vipRank or \Local may be preferred, whereas \abc and \Gmax provide robust protection against false discoveries when precision is paramount.

Several avenues for future work remain. A deeper theoretical understanding of MPVIP behavior under posterior concentration could help diagnose and correct for DART's FP inflation in high-SNR settings. Furthermore, extensions of VC-based measures to other ensemble methods or multimodal data settings present exciting opportunities for broader application.

\newpage

\vspace{\baselineskip}
\noindent \textbf{Supplementary Materials:} 
The Supplementary Materials, appended after the references, contains the pseudo-code of the VC-measure (Section~\ref{sec:pseudo}), a comparison of BART and DART under the VC-measure (Section~\ref{sec:BART_vs_DART_vc}), and a comparison of VC- and VIP-measure (Section~\ref{sec:vc_vs_vip}). Code for reproducing experiments and figures of this paper is available at \href{https://github.com/mattsheng/bayesVC}{github.com/mattsheng/bayesVC}.

\bibliography{ref} 


\pagebreak
\begin{center}
\textbf{\large Supplementary Material for ``Posterior Summarization for Variable Selection in Bayesian Tree Ensembles''}
\end{center}
\setcounter{equation}{0}
\setcounter{figure}{0}
\setcounter{table}{0}
\setcounter{page}{1}
\setcounter{section}{0}
\makeatletter
\renewcommand{\theequation}{S\arabic{equation}}
\renewcommand{\thefigure}{S\arabic{figure}}
\renewcommand{\bibnumfmt}[1]{[S#1]}
\renewcommand{\citenumfont}[1]{S#1}
\renewcommand{\thesection}{\Alph{section}} 
\renewcommand{\thetable}{S\arabic{table}}

\section{Pseudo Code for VC-measure}\label{sec:pseudo}
\begin{figure}[h!]
    \centering
    \begin{minipage}{0.85\linewidth}
    \begin{algorithm}[H]
    \caption{VC-measure}\label{alg:vc-measure}
    \begin{algorithmic}
    \Require data $(y, \bX)$, number of replications $\Lrep$
    \Ensure selected predictors $\bX_{\widehat{\mathcal{S}}}$
    \State Initialize $s \gets $ $\Lrep$ random integers 
    \For{$\ell = 1,\ldots,\Lrep$} 
        \State Train a Bayesian forest model on $(y, \bX)$ using random seed $s[\ell]$
        \State Compute VC from model $\ell$: $\bm{c}_{(\ell)} \gets (c_{1,\ell}, \ldots, c_{p,\ell}) \in \mathbb{R}^p$
        \State Rank $\bm{c}_{(\ell)}$ in descending order: $(R(c_{1,\ell}), \ldots, R(c_{p,\ell})) \in \{1,\ldots,p\}^p$
    \EndFor
    \For{$j = 1, \ldots, p$}
        \State $\bar{c}_j \gets \frac{1}{\Lrep}\sum_{\ell=1}^\Lrep c_{j,\ell}$
        \State $\overline{R}_j^c \gets \frac{1}{\Lrep}\sum_{\ell=1}^\Lrep R(c_{j,\ell})$
        \State $\mathcal{Q}_{0.25}(\bm{c}_j) \gets 25\%\text{ Quantile of } (c_{j,1}, \ldots, c_{j,\Lrep})$
        \State $\mathcal{Q}_{0.75}(\bm{R}_j^c) \gets 75\%\text{ Quantile of } (R(c_{j,1}),\ldots,R(c_{j,\Lrep}))$
        \State $\bm{Z}_j \gets \left(\bar{c}_j, \mathcal{Q}_{0.25}(\bm{c}_j), \overline{R}_j^c, \mathcal{Q}_{0.75}(\bm{R}_j^c)\right) \in \mathbb{R}^4$
    \EndFor
    \State $\bm{Z} \in \mathbb{R}^{p \times 4} \gets$ stacking $\bm{Z}_j$ row-wise
    \State $\widetilde{\bm{Z}} \gets$ \texttt{log1p} transformation (i.e., $\log (1 + x)$) followed by column standardization
    \State Perform HAC on $\widetilde{\bm{Z}}$, cut dendrogram to form 2 clusters, and compute their average VC
    \State Return $\widehat{\mathcal{S}} \gets$ predictor indices of the high VC cluster
    \end{algorithmic}
    \end{algorithm}
    \end{minipage}
\end{figure}

\section{Comparison of BART and DART Using VC-measure}\label{sec:BART_vs_DART_vc}

In this section, we compare BART and DART when paired with the same variable importance measure, VC-measure, to assess whether the sparsity-inducing Dirichlet prior in DART improves variable selection accuracy. For each of the 3,600 simulation settings in Table~\ref{tab:setting}, each with 10 replications, we independently train $\Lrep = 10$ BART and DART models and compute their respective VC-measures.  

\begin{figure}[t]
    \centering
    \includegraphics[width=\linewidth]{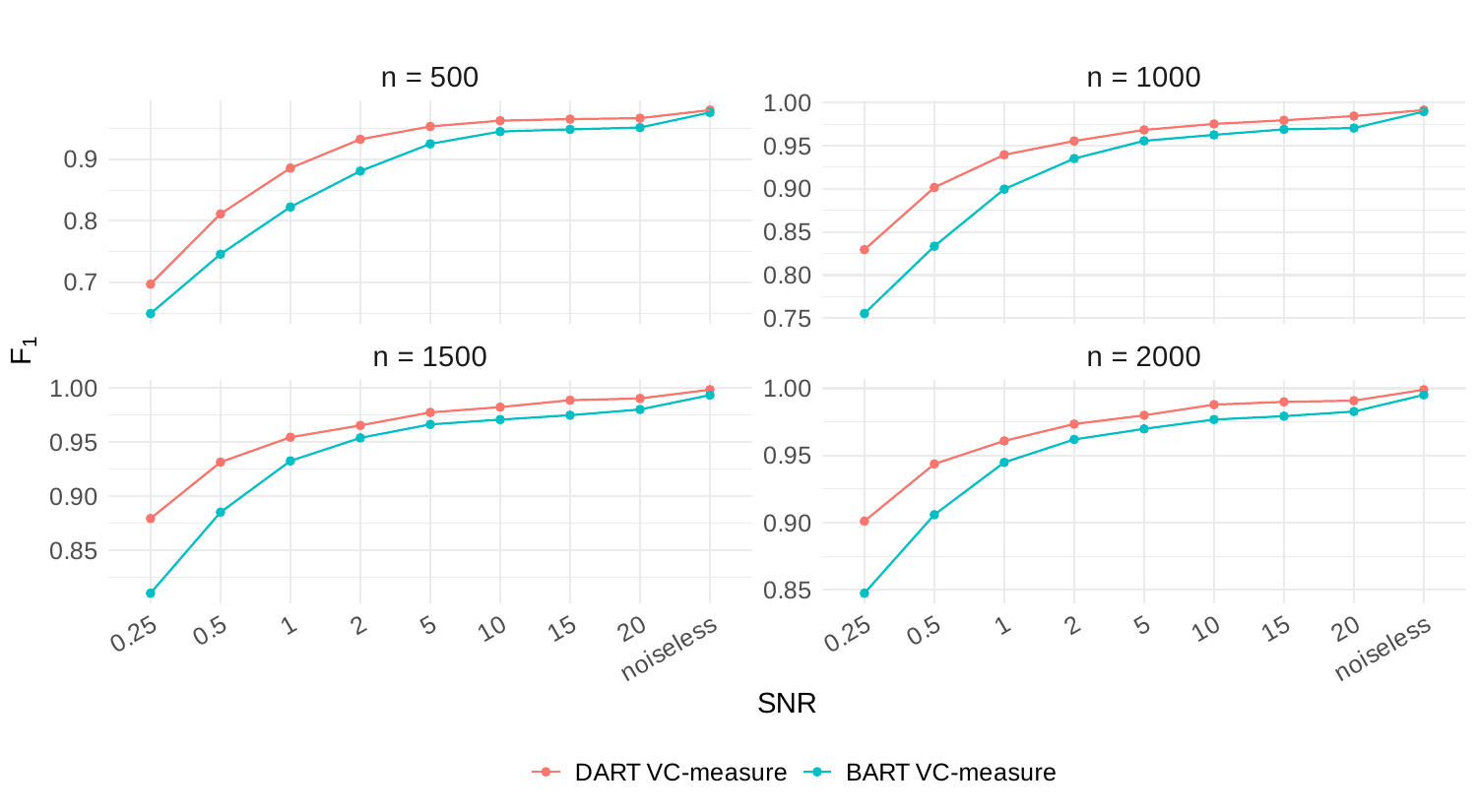}
    \caption{$F_1$ score of DART vs BART using VC-measure variable importance measure. Points indicate the average $F_1$ scores over 100 Feynman equations, each with 10 replicates.} %
    \label{fig:DART_vs_BART}
\end{figure}

\begin{figure}[tbh!]
    \centering
    \includegraphics[width=0.85\linewidth]{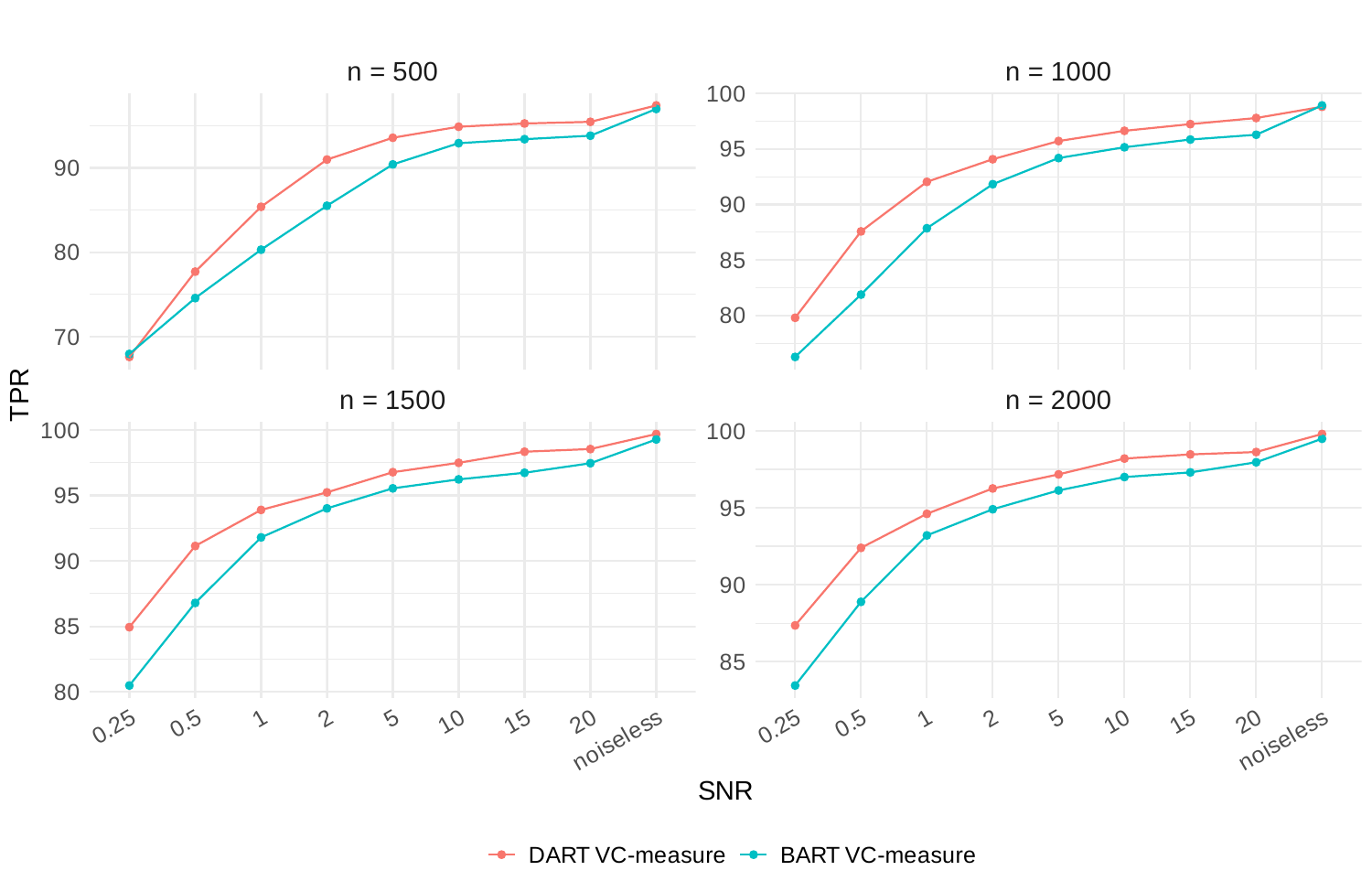}
    \caption{TPR of DART vs BART using VC-measure variable importance measure. Points indicate the average TPR over 100 Feynman equations, each with 10 replicates.}
    \label{fig:DART_vs_BART_TPR}
\end{figure}

\begin{figure}[h!]
    \centering
    \includegraphics[width=0.85\linewidth]{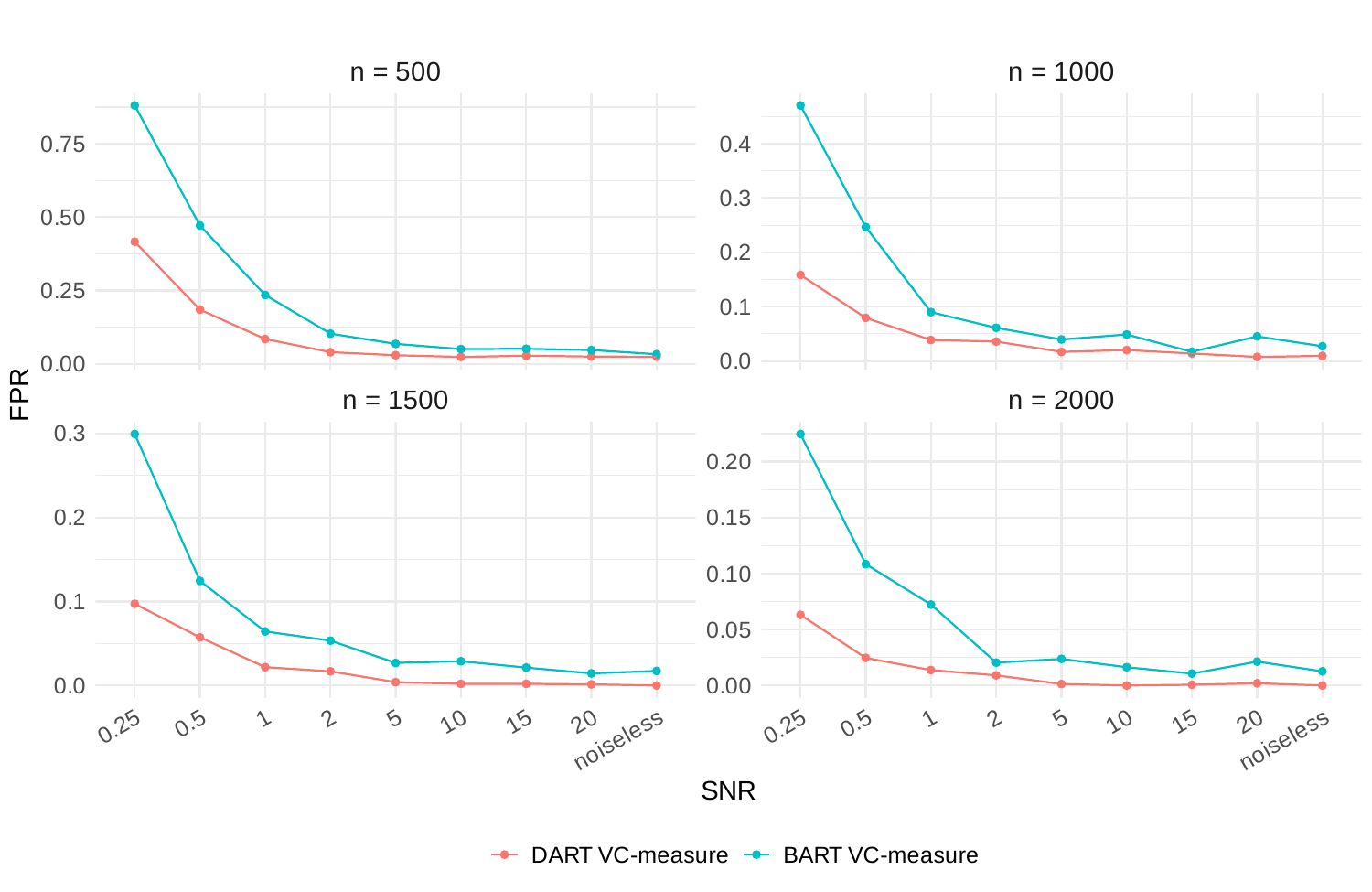}
    \caption{FPR of DART vs BART using VC-measure variable importance measure. Points indicate the average FPR over 100 Feynman equations, each with 10 replicates.}
    \label{fig:DART_vs_BART_FPR}
\end{figure}

Since both methods share the same importance measure, any difference in selection accuracy reflects differences in posterior quality. As shown in Figure~\ref{fig:DART_vs_BART}, \vcMeasure consistently outperforms BART VC-measure in $F_1$ score across all combinations of $(n,\text{SNR})$. This suggests that posterior samples drawn under the DART prior provide a clearer separation boundary in VC-measure. Similar improvements are observed in TPR and FPR (Figures~\ref{fig:DART_vs_BART_TPR} and \ref{fig:DART_vs_BART_FPR}): \vcMeasure attains a higher TPR than BART VC-measure in 34 of the 36 $(n, \text{SNR})$ settings and a lower FPR in all 36. However, this performance gain comes with a modest computational cost, as illustrated in Table~\ref{tab:runtime_BART_vs_DART}, which is attributed to the additional Gibbs sampling steps for the variable splitting probabilities $\bm{s}$ introduced by the Dirichlet prior in DART.

\begin{table*}[b]
\caption{Average runtime (in seconds) across various sample sizes $n$.} \label{tab:runtime_BART_vs_DART}
\vskip 0.1in
\begin{center}
\begin{small}
\begin{sc}
\begin{tabular}{lllll}
\toprule
                & $n=500$ & $n=1000$ & $n=1500$ & $n=2000$ \\
\midrule
BART VC-measure    & 55.59   & 93.19    & 136.73   & 183.35 \\
\vcMeasure         & 105.76  & 155.59   & 202.08   & 247.75 \\
\bottomrule
\end{tabular}
\end{sc}
\end{small}
\end{center}
\end{table*}

\begin{figure}[tbh!]
    \centering
    \includegraphics[width=\linewidth]{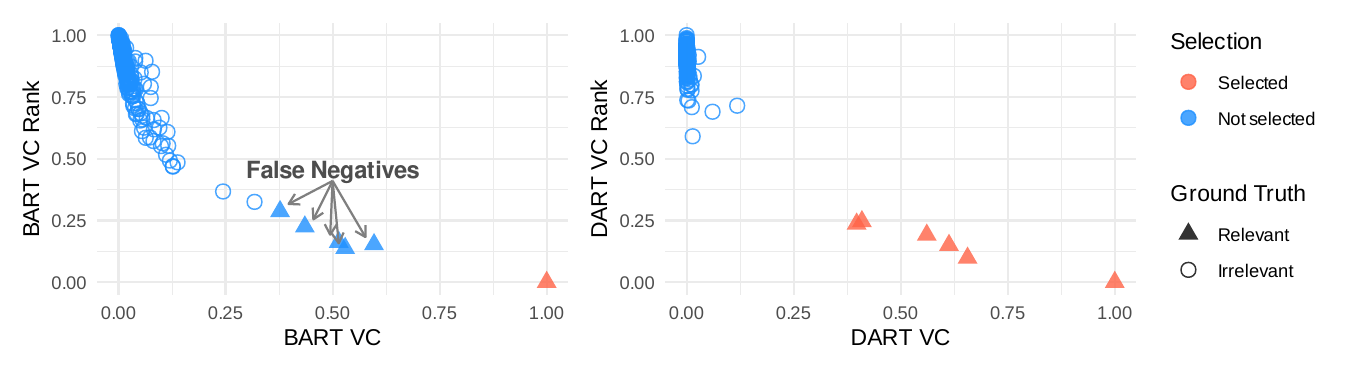}
    \caption{Scatterplot of BART VC (on the left) and DART VC (on the right). Model is trained on the Feynman II-11-17 equation: $y = x_1\left(1+\frac{x_5x_6\cos(x_4)}{x_2x_3} \right) + \varepsilon$, with $n=1000$, $p_0=6$, $p=306$, and $\text{SNR}=1$. Each point corresponds to one predictor. Ground truth predictor type is distinguished by a triangle for relevant predictors $(x_1,\ldots,x_6)$ and an empty circle for irrelevant predictors $(x_7,\ldots,x_{306})$. Selection decision is distinguished by color, where red indicates ``Selected'' and blue indicates ``Not Selected.''}
    \label{fig:scatter_BART_vs_DART_VC}
\end{figure}

To visualize the impact of the Dirichlet prior, we train BART and DART on the Feynman II-11-17 equation $y = x_1\left(1 + \frac{x_5x_6\cos(x_4)}{x_2x_3} \right) + \varepsilon$, and compare signal separability based on VC and VC Rank. To better visualize the differences, all measures are $\log(1+x)$-transformed, followed by a min-max normalization to a common scale of $[0,1]$.

As shown in Figure~\ref{fig:scatter_BART_vs_DART_VC}, DART VC-measure exhibits a clearer separation between $\Xrel$ and $\Xirr$ along both axes (VC on the $x$-axis and VC Rank on the $y$-axis). In particular, BART assigns higher VCs to $\Xirr$ (denoted by empty circles) than DART does, indicating that BART splits on these irrelevant variables more frequently. This reflects BART's weaker regularization and inability to suppress irrelevant splits. In contrast, DART's VC of $\Xirr$ tightly around 0, confirming that the Dirichlet prior effectively penalizes their inclusion in the tree ensemble. This improved regularization translates to enhanced signal-noise separability and leads to better selection accuracy. 

These results reinforce the importance of incorporating sparsity-inducing priors in Bayesian tree ensembles for high-dimensional variable selection. While VC-measure alone offers substantial gains across Bayesian tree ensemble models, its combination with the DART prior consistently yields superior selection accuracy. In fact, DART VC-measure consistently outperforms all other methods as detailed in Section~\ref{sec:comparison}. The Dirichlet prior in DART not only improves posterior concentration around signal variables but also actively discourages spurious splits on noise, thereby enhancing the interpretability and stability of variable rankings. Together, these findings highlight that both the design of the importance measure and the choice of prior play critical and complementary roles in effective nonparametric variable selection.

\section{Comparison of VC-measure and VIP-measure} \label{sec:vc_vs_vip}
\begin{figure}[t]
    \centering
    \includegraphics[width=\linewidth]{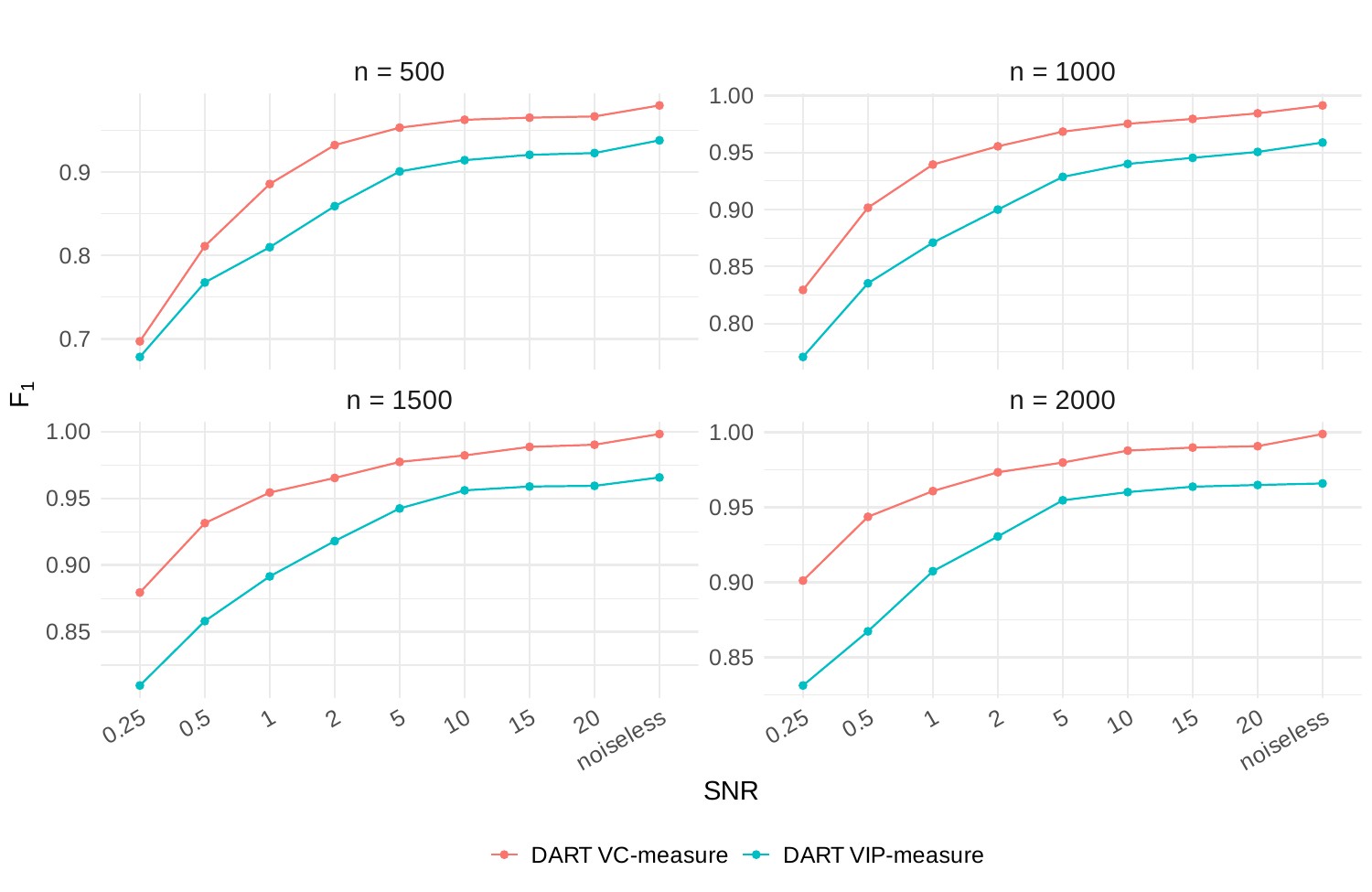}
    \caption{$F_1$ score of \vcMeasure vs DART VIP-measure. Points indicate the average $F_1$ scores over 100 Feynman equations, each with 10 replicates.}
    \label{fig:VC_vs_VIP}
\end{figure}
\begin{figure}[h!]
    \centering
    \includegraphics[width=\linewidth]{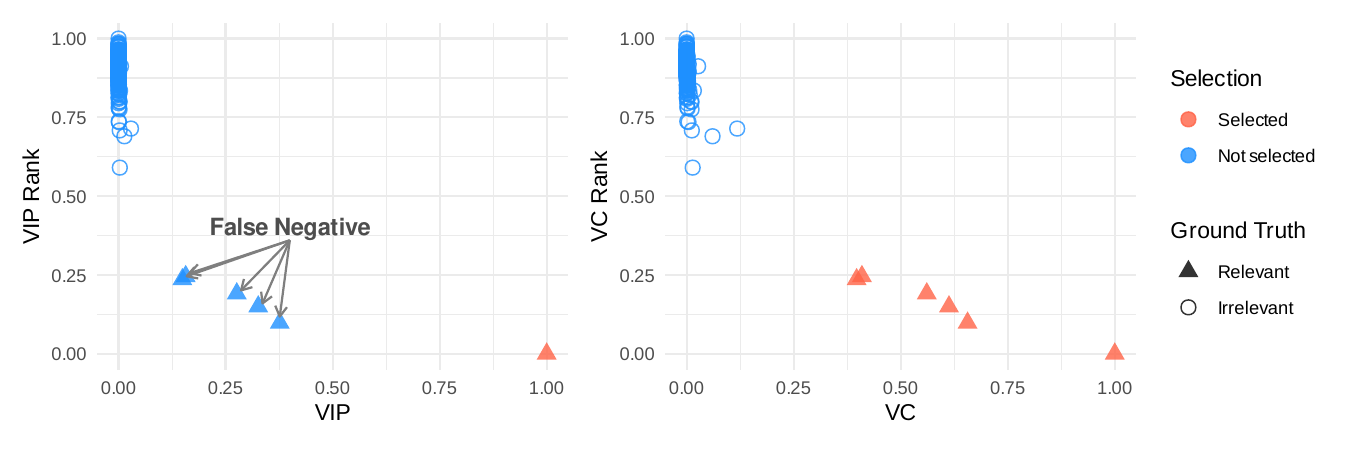}
    \caption{Scatterplot of VIP-measure (on the left) and VC-measure (on the right) variable importance measures calculated from identical DART posterior samples. Model is trained on the Feynman II-11-17 equation: $y = x_1\left(1+\frac{x_5x_6\cos(x_4)}{x_2x_3} \right) + \varepsilon$, with $n=1000$, $p_0=6$, $p=306$, and $\text{SNR}=1$. Each point corresponds to one predictor. Ground truth predictor type is distinguished by a triangle for relevant predictors $(x_1,\ldots,x_6)$ and an empty circle for irrelevant predictors $(x_7,\ldots,x_{306})$. Selection decision is distinguished by color, where red indicates ``Selected'' and blue indicates ``Not Selected.''}
    \label{fig:scatter_VC_vs_VIP}
\end{figure}

Although VC can be obtained as a special case of VIP, by setting the per-draw total variable count $c_{\cdot,k}$ to 1 (or a constant) for all $k$, this relationship does not imply that VIP matches VC in performance. In practice, variation in $c_{\cdot,k}$ directly affects both the scale and stability of the resulting importance measures. The VIP normalization can inadvertently shrink the variable count $c_{j,k}$ of relevant predictors $\Xrel$ when the denominator $c_{\cdot,k}$ is inflated by frequent splits on $\Xirr$, while also amplifying variability across posterior draws. In contrast, VC omits this normalization, preserving the magnitude of counts for informative predictors and reducing instability introduced by irrelevant ones. In this section, we directly compare VC-measure and VIP-measure to assess their relative effectiveness in variable selection.

For each of the 3,600 simulation settings in Table~\ref{tab:setting}, each with 10 replications, we train $\Lrep = 10$ DART models and compute both VC- and VIP-measure using the same posterior samples. This ensures that any difference in selection accuracy arises solely from the importance measure itself. HAC is then applied separately to VC- and VIP-measure, yielding two sets of selected variables per setting. 

\begin{figure}[tbh!]
    \centering
    \includegraphics[width=0.85\linewidth]{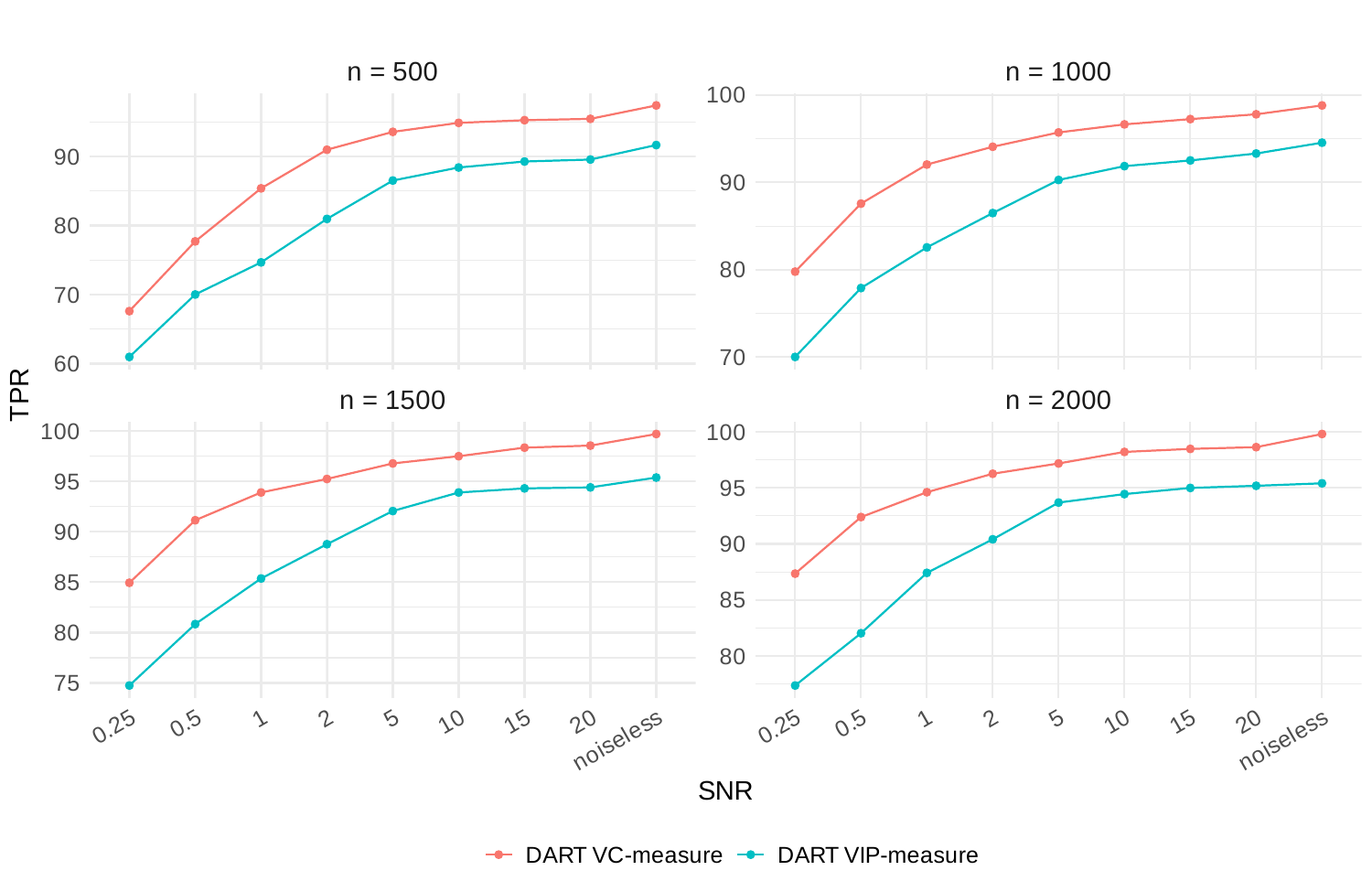}
    \caption{TPR of DART VC-measure vs DART VIP-measure. Points indicate the average TPR over 100 Feynman equations, each with 10 replicates.}
    \label{fig:VC_vs_VIP_tpr}
\end{figure}

\begin{figure}[tbh!]
    \centering
    \includegraphics[width=0.85\linewidth]{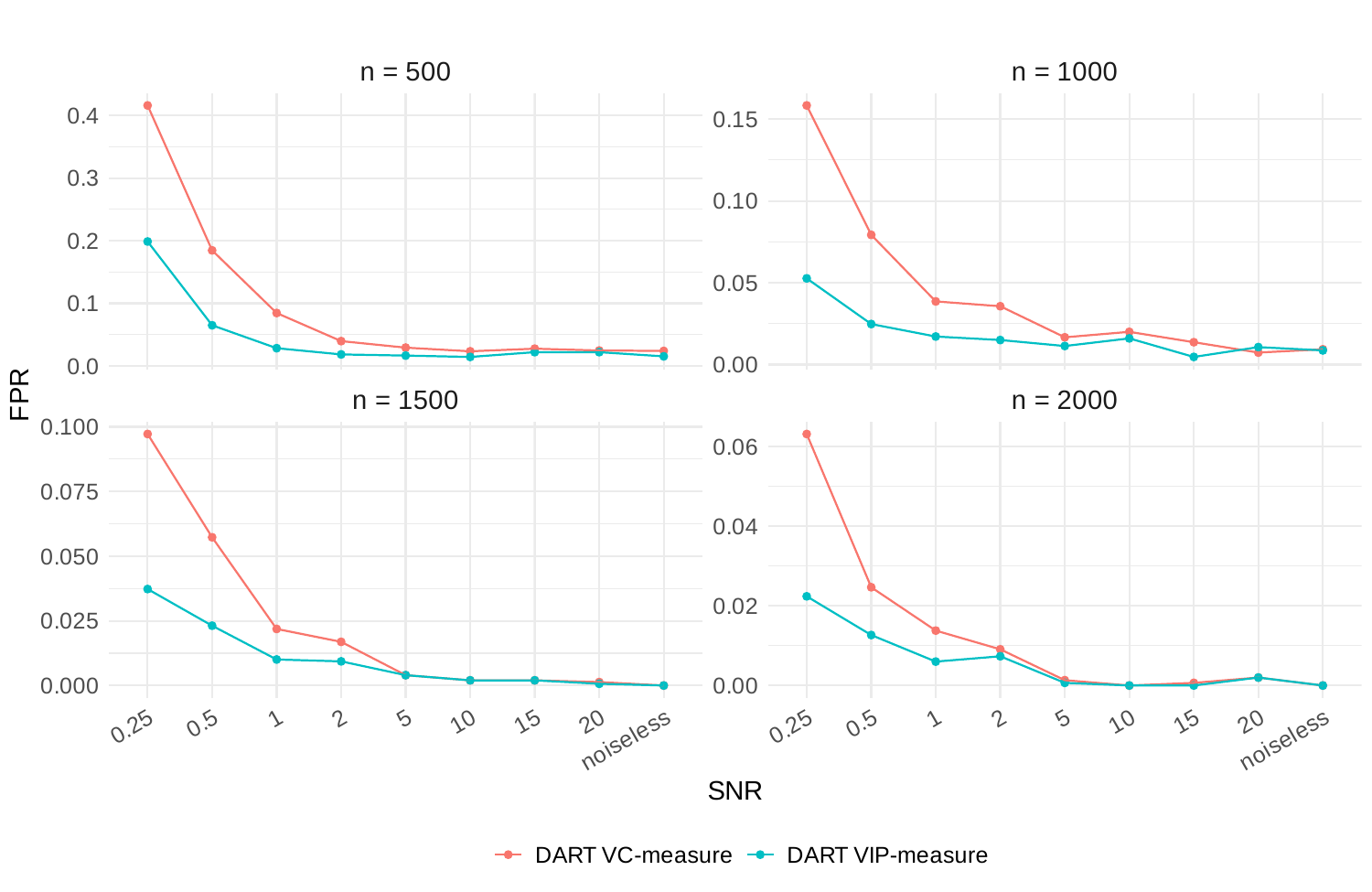}
    \caption{FPR of DART VC-measure vs DART VIP-measure. Points indicate the average FPR over 100 Feynman equations, each with 10 replicates.}
    \label{fig:VC_vs_VIP_fpr}
\end{figure}

Figure~\ref{fig:VC_vs_VIP} shows that \vcMeasure consistently outperforms its VIP counterpart in terms of $F_1$ score across all combinations of $n$ and SNR. Moreover, in all but the four noisiest settings with $n \leq 1000$ ($\text{SNR} \leq 0.5$), the performance gain here is more substantial than that of switching from BART to DART (Figure~\ref{fig:DART_vs_BART}), suggesting that a better importance measure may play a more important role than a sparsity-inducing prior.
The improvement in $F_1$ is driven by gains in TPR of 3--11 percentage points in every setting (Figure~\ref{fig:VC_vs_VIP_tpr}), at the cost of a slightly higher FPR in most settings (Figure~\ref{fig:VC_vs_VIP_fpr}); both FPRs stay below 0.5\%, i.e., fewer than one false positive per run on average. In unreported results, we repeat the same experiment using posterior samples from BART. The results are analogous: BART VC-measure consistently surpasses BART VIP-measure in $F_1$ and TPR, indicating that the advantage of VC-measure over VIP-measure is not specific to DART. %

We further illustrate this point using the Feynman II-11-17 equation with $n=1000$, $p_0=6$, $p=306$, and $\text{SNR=1}$. We train $\Lrep =10$ DART models and compare the resulting VIP-measure (left panel) and VC-measure (right panel) in Figure~\ref{fig:scatter_VC_vs_VIP}. Variable types are indicated by shapes: triangles for relevant predictors $\Xrel=(x_1,\ldots,x_6)$ and empty circles for irrelevant predictors $\Xirr = (x_7,\ldots,x_{306})$. Selection status is color-coded, where red indicates ``Selected'' and blue indicates ``Not Selected.'' While VIP Rank and VC Rank ($y$-axes) display similar distributions, VIP ($x$-axis, left panel) places five of the six relevant predictors closer to the blue ``Not Selected'' cluster, causing them to be missed by HAC. In contrast, these same predictors are positioned farther from the ``Not Selected'' cluster, demonstrating VC-measure's superior signal-noise separability.

In summary, although VIP and VC share the same conceptual foundation, the absence of normalization in VC avoids the shrinkage of relevant predictors and reduces the noise-induced variability that can occur in VIP. This leads to a clearer and more stable separation between $\Xrel$ and $\Xirr$, producing consistent gains in $F_1$ and TPR at a negligible cost in FPR. These results confirm that VC-measure is not only a computationally simpler alternative to VIP, but also a more accurate one for Bayesian tree ensemble variable selection.

\end{document}